\documentclass[letterpaper,prc,superscriptaddress,showpacs,twocolumn,amssymb,amsmath,amsfonts,aps]{revtex4}
\usepackage[pdftex]{graphicx}
\usepackage{dcolumn}  
\usepackage{longtable}
\usepackage{bm}       
\usepackage{subfigure}
\usepackage{rotating}
\usepackage{epstopdf}
\newcommand\captionof[1]{\def\@captype{#1}\caption}
\begin{document}
%
%
%
%
\newcommand*{\CMU}{Carnegie Mellon University, Pittsburgh, Pennsylvania 15213}
\newcommand*{\CMUindex}{1}
\affiliation{\CMU}
\newcommand*{\ANL}{Argonne National Laboratory, Argonne, Illinois 60441}
\newcommand*{\ANLindex}{2}
\affiliation{\ANL}
\newcommand*{\ASU}{Arizona State University, Tempe, Arizona 85287-1504}
\newcommand*{\ASUindex}{3}
\affiliation{\ASU}
\newcommand*{\CSU}{California State University, Dominguez Hills, Carson, CA 90747}
\newcommand*{\CSUindex}{4}
\affiliation{\CSU}
\newcommand*{\Canisius}{Canisius College, Buffalo, NY 14208}
\newcommand*{\Canisiusindex}{5}
\affiliation{\Canisius}
\newcommand*{\CUA}{Catholic University of America, Washington, D.C. 20064}
\newcommand*{\CUAindex}{6}
\affiliation{\CUA}
\newcommand*{\SACLAY}{CEA, Centre de Saclay, Irfu/Service de Physique Nucl\'eaire, 91191 Gif-sur-Yvette, France}
\newcommand*{\SACLAYindex}{7}
\affiliation{\SACLAY}
\newcommand*{\CNU}{Christopher Newport University, Newport News, Virginia 23606}
\newcommand*{\CNUindex}{8}
\affiliation{\CNU}
\newcommand*{\UCONN}{University of Connecticut, Storrs, Connecticut 06269}
\newcommand*{\UCONNindex}{9}
\affiliation{\UCONN}
\newcommand*{\ECOSSEE}{Edinburgh University, Edinburgh EH9 3JZ, United Kingdom}
\newcommand*{\ECOSSEEindex}{10}
\affiliation{\ECOSSEE}
\newcommand*{\FU}{Fairfield University, Fairfield CT 06824}
\newcommand*{\FUindex}{11}
\affiliation{\FU}
\newcommand*{\FIU}{Florida International University, Miami, Florida 33199}
\newcommand*{\FIUindex}{12}
\affiliation{\FIU}
\newcommand*{\FSU}{Florida State University, Tallahassee, Florida 32306}
\newcommand*{\FSUindex}{13}
\affiliation{\FSU}
\newcommand*{\GWU}{The George Washington University, Washington, DC 20052}
\newcommand*{\GWUindex}{14}
\affiliation{\GWU}
\newcommand*{\ECOSSEG}{University of Glasgow, Glasgow G12 8QQ, United Kingdom}
\newcommand*{\ECOSSEGindex}{15}
\affiliation{\ECOSSEG}
\newcommand*{\ISU}{Idaho State University, Pocatello, Idaho 83209}
\newcommand*{\ISUindex}{16}
\affiliation{\ISU}
\newcommand*{\INFNFR}{INFN, Laboratori Nazionali di Frascati, 00044 Frascati, Italy}
\newcommand*{\INFNFRindex}{17}
\affiliation{\INFNFR}
\newcommand*{\INFNGE}{INFN, Sezione di Genova, 16146 Genova, Italy}
\newcommand*{\INFNGEindex}{18}
\affiliation{\INFNGE}
\newcommand*{\INFNRO}{INFN, Sezione di Roma Tor Vergata, 00133 Rome, Italy}
\newcommand*{\INFNROindex}{19}
\affiliation{\INFNRO}
\newcommand*{\ORSAY}{Institut de Physique Nucl\'eaire ORSAY, Orsay, France}
\newcommand*{\ORSAYindex}{20}
\affiliation{\ORSAY}
\newcommand*{\ITEP}{Institute of Theoretical and Experimental Physics, Moscow, 117259, Russia}
\newcommand*{\ITEPindex}{21}
\affiliation{\ITEP}
\newcommand*{\JMU}{James Madison University, Harrisonburg, Virginia 22807}
\newcommand*{\JMUindex}{22}
\affiliation{\JMU}
\newcommand*{\KYUNGPOOK}{Kyungpook National University, Daegu 702-701, Republic of Korea}
\newcommand*{\KYUNGPOOKindex}{23}
\affiliation{\KYUNGPOOK}
\newcommand*{\UNH}{University of New Hampshire, Durham, New Hampshire 03824-3568}
\newcommand*{\UNHindex}{24}
\affiliation{\UNH}
\newcommand*{\NSU}{Norfolk State University, Norfolk, Virginia 23504}
\newcommand*{\NSUindex}{25}
\affiliation{\NSU}
\newcommand*{\OHIOU}{Ohio University, Athens, Ohio  45701}
\newcommand*{\OHIOUindex}{26}
\affiliation{\OHIOU}
\newcommand*{\ODU}{Old Dominion University, Norfolk, Virginia 23529}
\newcommand*{\ODUindex}{27}
\affiliation{\ODU}
\newcommand*{\RPI}{Rensselaer Polytechnic Institute, Troy, New York 12180-3590}
\newcommand*{\RPIindex}{28}
\affiliation{\RPI}
\newcommand*{\URICH}{University of Richmond, Richmond, Virginia 23173}
\newcommand*{\URICHindex}{29}
\affiliation{\URICH}
\newcommand*{\ROMAII}{Universita' di Roma Tor Vergata, 00133 Rome Italy}
\newcommand*{\ROMAIIindex}{30}
\affiliation{\ROMAII}
\newcommand*{\MOSCOW}{Skobeltsyn Nuclear Physics Institute, Skobeltsyn Nuclear Physics Institute, 119899 Moscow, Russia}
\newcommand*{\MOSCOWindex}{31}
\affiliation{\MOSCOW}
\newcommand*{\SCAROLINA}{University of South Carolina, Columbia, South Carolina 29208}
\newcommand*{\SCAROLINAindex}{32}
\affiliation{\SCAROLINA}
\newcommand*{\JLAB}{Thomas Jefferson National Accelerator Facility, Newport News, Virginia 23606}
\newcommand*{\JLABindex}{33}
\affiliation{\JLAB}
\newcommand*{\UNIONC}{Union College, Schenectady, NY 12308}
\newcommand*{\UNIONCindex}{34}
\affiliation{\UNIONC}
\newcommand*{\UTFSM}{Universidad T\'{e}cnica Federico Santa Mar\'{i}a, Casilla 110-V Valpara\'{i}so, Chile}
\newcommand*{\UTFSMindex}{35}
\affiliation{\UTFSM}
\newcommand*{\VIRGINIA}{University of Virginia, Charlottesville, Virginia 22901}
\newcommand*{\VIRGINIAindex}{36}
\affiliation{\VIRGINIA}
\newcommand*{\WM}{College of William and Mary, Williamsburg, Virginia 23187-8795}
\newcommand*{\WMindex}{37}
\affiliation{\WM}
\newcommand*{\YEREVAN}{Yerevan Physics Institute, 375036 Yerevan, Armenia}
\newcommand*{\YEREVANindex}{38}
\affiliation{\YEREVAN}

\newcommand*{\NOWIMPERIAL}{Imperial College London, London SW7 2AZ, UK.} 
\newcommand*{\NOWSTANFORD}{Stanford University, Stanford, CA 94305, USA.}
\newcommand*{\NOWLPSC}{LPSC-Grenoble, France.}
\newcommand*{\NOWJLAB}{Thomas Jefferson National Accelerator Facility, Newport News, Virginia 23606, USA.}
\newcommand*{\NOWLANL}{Los Alamos National Laborotory, New Mexico, NM, USA.}
\newcommand*{\NOWMINN}{University of Minnesota, Minneapolis, MN 55455, USA.}
\newcommand*{\NOWECOSSEE}{Edinburgh University, Edinburgh EH9 3JZ, UK.}
\newcommand*{\NOWWM}{College of William and Mary, Williamsburg, Virginia 23187, USA.}
\newcommand*{\NOWGW}{The George Washington University, Washington, DC 20052, USA.}

\author{M.~Williams}
\altaffiliation[Current address: ]{\NOWIMPERIAL}
\affiliation{\CMU}
\author {D.~Applegate}
\altaffiliation[Current address: ]{\NOWSTANFORD}
\affiliation{\CMU}
\author {M.~Bellis} 
\affiliation{\CMU}
\author {C.A.~Meyer}
\affiliation{\CMU}
\author {K. P. ~Adhikari} 
\affiliation{\ODU}
\author {M.~Anghinolfi} 
\affiliation{\INFNGE}
\author {H.~Baghdasaryan} 
\affiliation{\VIRGINIA}
\affiliation{\ODU}
\author {J.~Ball} 
\affiliation{\SACLAY}
\author {M.~Battaglieri} 
\affiliation{\INFNGE}
\author {I.~Bedlinskiy} 
\affiliation{\ITEP}
% ...

\collaboration{The CLAS Collaboration}
\noaffiliation

%
 
%
%
%
%
%
%


\date{\today}

\title{Differential cross sections and spin density matrix elements for 
  the reaction $\gamma p \rightarrow p \omega$}
%
%
\begin{abstract} 
High-statistics differential cross sections and spin density matrix elements
for the reaction $\gamma p \rightarrow p \omega$ have been measured using the
CEBAF large acceptance spectrometer (CLAS) at Jefferson Lab for center-of-mass 
(c.m.) energies from threshold up to
2.84~GeV. Results are reported in 112 10-MeV wide c.m. energy bins, each 
subdivided into $\cos{\theta_{c.m.}^{\omega}}$ bins of width 0.1. 
These are the most precise and extensive $\omega$ photoproduction measurements
to date.
A number of prominent structures are clearly present in the data.
Many of these have not previously been observed due to limited statistics in 
earlier measurements.
\end{abstract}
\maketitle
\section{\label{section:intro}INTRODUCTION}

Studying low-energy $\omega$ photoproduction presents an
interesting opportunity to search for new baryon resonances. 
Previous experiments have produced cross-section measurements with relatively
high precision at most production angles; however,  precise spin density 
matrix elements have only been measured at very forward  angles
\cite{ballam-1973,clift-1977,barber-1984,battaglieri-2003,saphir-2003}. 
Theoretical interpretation of these data indicate strong $t$-channel 
contributions from both $\pi^0$ and Pomeron exchange, while the backward peak 
in the cross section has been interpreted as evidence of nucleon $u$-channel 
contributions~\cite{laget-2000,laget-2002,oh-2001,sibirtsev-2002}.
Several attempts have been made to extract resonant contributions that have
obtained conflicting 
results~\cite{paris-09,zhao-2001,titov,oh-2001,penner-2002,penner-2005}. 
Precise polarization information is needed in order to
place stringent constraints on the physics interpretation of  $\omega$ 
photoproduction data.

The impact of polarization information can be seen by comparing the
partial wave analysis 
results obtained using only cross-section data~\cite{penner-2002} 
to those that also included the low-precision polarization results from 
SAPHIR~\cite{penner-2005}. The former found that at threshold the dominant 
$s$-channel contributions are from the $P_{13}(1720)$ and $F_{15}(1680)$, while
the latter found that the $D_{15}(1675)$ and $F_{15}(1680)$ are dominant in
this region.  
Including polarization information, even with very limited precision, 
provided strong additional constraints on the interpretation of the data. 
Thus, obtaining high-precision polarization results is a vital step towards 
understanding baryon resonance contributions to $\omega$ photoproduction.  
 
Beyond this, quark model calculations of baryon decays~\cite{capstick-1993} 
predict that a number of the so-called \emph{missing} baryons should couple to 
$\omega N$ final states. In particular, in the above model, nearly all of the 
missing positive parity $N^*$ states are expected to have non-negligible 
couplings to $\omega N$. Thus, good data on  $\omega$ photoproduction coupled 
with a partial wave analysis 
(PWA) could provide important new information on light-quark baryons. 

The data presented here are part of a larger program to simultaneously measure 
photoproduction of mesons and then carry out partial wave analyses on the 
resulting data. This article presents  
differential cross section and spin density matrix element measurements
for $\omega$ photoproduction. 
In a companion article published concurrent to this one~\cite{williams-prd}, 
we present a detailed partial wave analysis of these data where clear 
$s$-channel resonance contributions are identified. 
A forthcoming article will discuss the impact on current theoretical models 
and coupled-channel analyses of these new precise 
measurements~\cite{williams-prl}.

\begin{figure}
  \centering
\subfigure[]{
  \label{confidence_level}
  \includegraphics[width=0.48\textwidth]{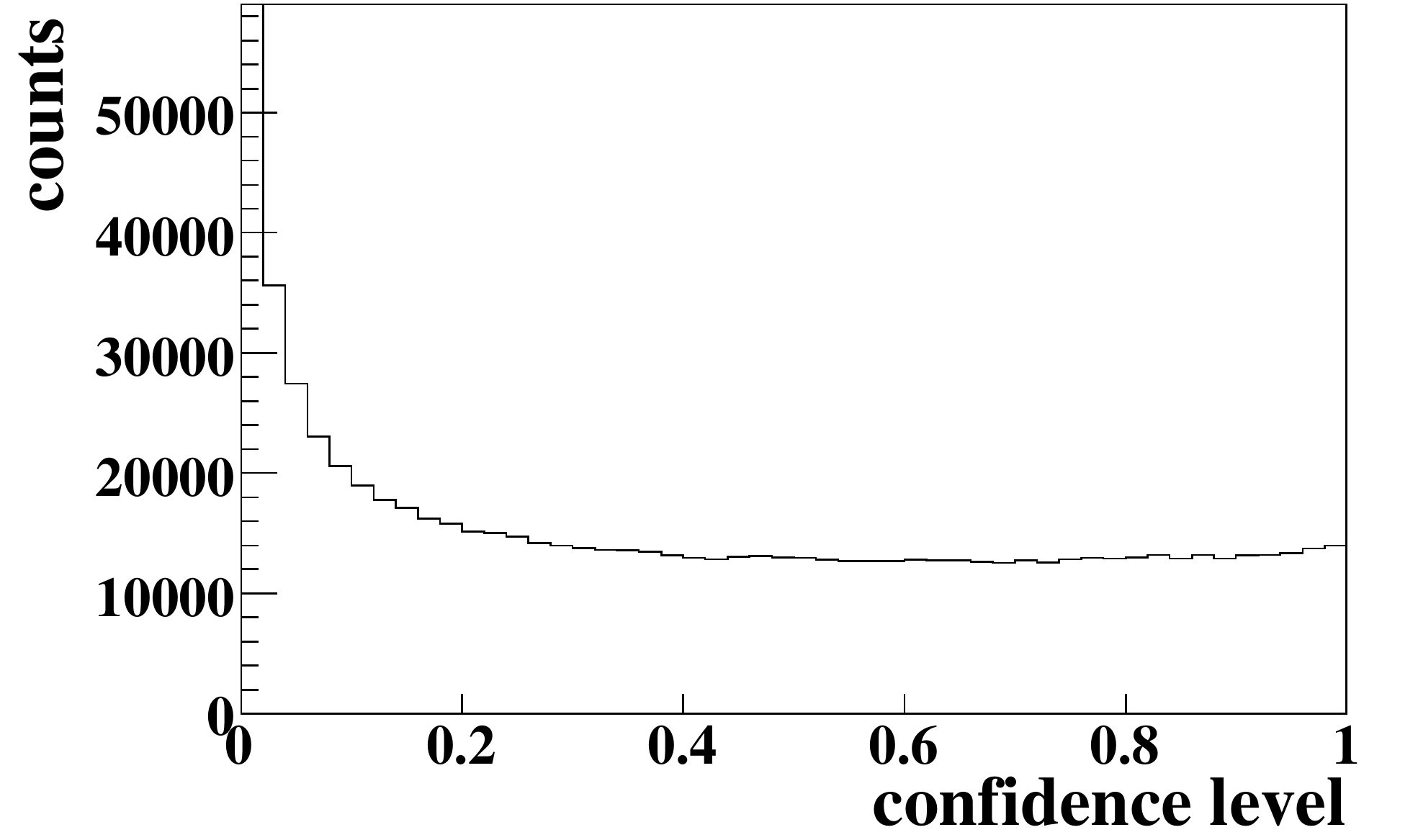}
} \\
\subfigure[]{
  \label{pull_dist}
  \includegraphics[width=0.48\textwidth]{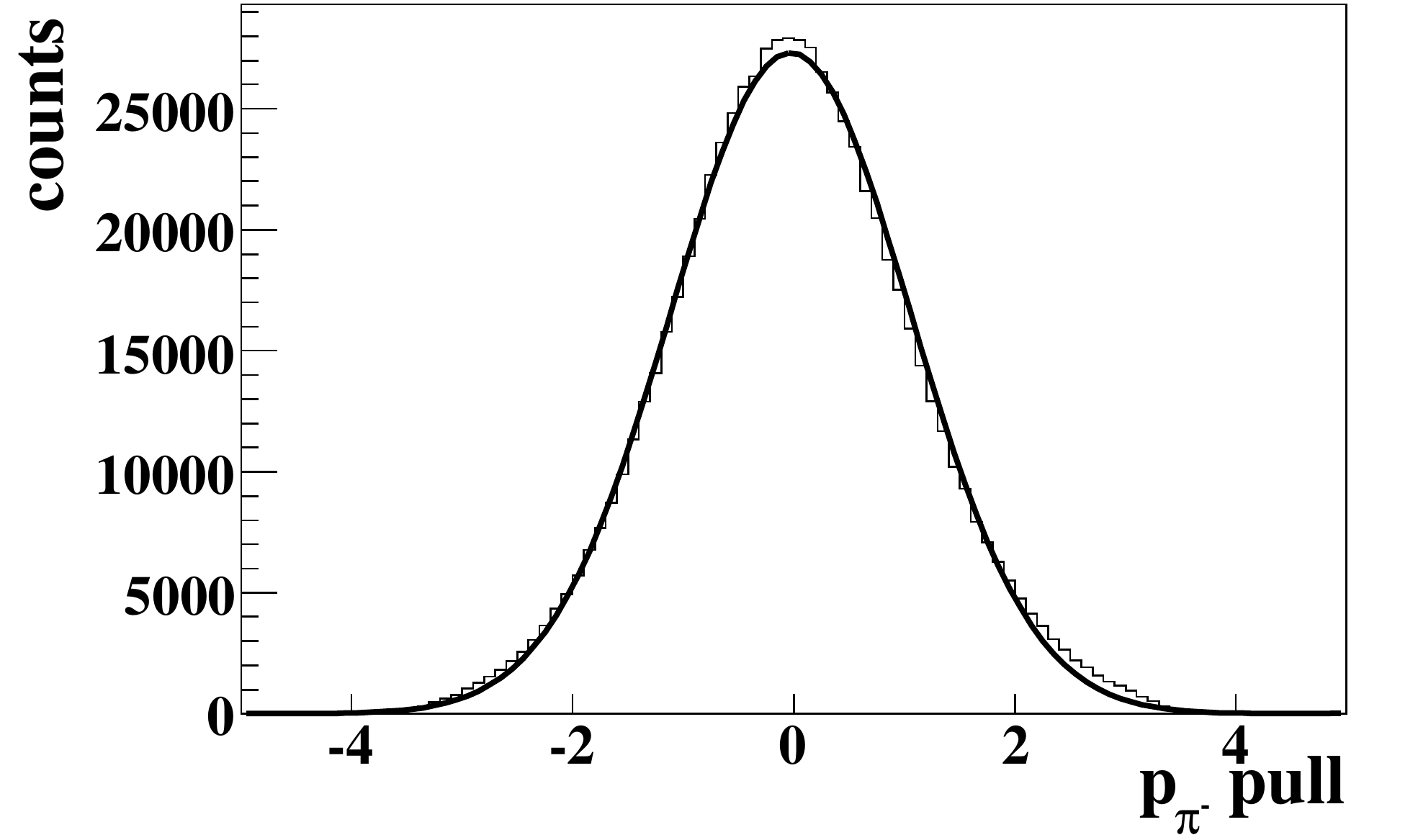}
}
\caption[]{\label{fig:kfit}
  (a) The confidence levels resulting from four-constraint kinematic fits 
  performed on a sample of events to the hypothesis
  $\gamma p \rightarrow p \pi^+ \pi^-$ integrated over all kinematics. 
  The ``peak'' near zero consists of events that do not match the hypothesis,
  along with poorly measured (due to multiple scattering, {\em etc}.) 
  signal events. Agreement with the ideal (flat)
  distribution for signal events is very good.
  (b) Example pull-distribution for the momentum of the $\pi^{-}$ from the same
  kinematic fits as in (a). 
  Only events with a confidence level larger than 1\% are shown. The line 
  represents a Gaussian fit to this distribution. For this event sample, the
  parameters obtained are $\mu = -0.029\pm0.001,\sigma = 1.086\pm0.001$ 
  (the uncertainties are purely statistical), which
  are in very good agreement with the ideal values $\mu = 0,\sigma = 1$.
  Both (a) and (b) are good indicators that the CLAS error matrix is well
  understood.
}
\end{figure}

\section{\label{section:setup}EXPERIMENTAL SETUP}

The data were obtained using the CEBAF large acceptance spectrometer (CLAS) 
housed in Hall B at the Thomas
Jefferson National Accelerator Facility. Real photons were produced via 
bremsstrahlung from a 4~GeV electron beam hitting a $10^{-4}$ radiation length
gold foil. The recoiling electrons were then analyzed using a dipole magnet
and scintillator hodoscopes in order to obtain, or ``tag'', the energy of the 
photons~\cite{sober} (the so-called photon tagger). 
The tagging range and energy resolution were $20\%-95\%$ and 
$0.1\%$ of the electron beam energy, respectively. The useful center-of-mass (c.m.) 
energy ($W$) range for this analysis was from $\omega$-photoproduction threshold at 
$W = 1.72$~GeV up to $2.84$~GeV. In this range, the data were analyzed in $10$-MeV 
wide $W$ bins.

The physics target, which was filled with liquid hydrogen, was a $40$-cm long
cylinder with a radius of $2$~cm. Continuous monitoring of the temperature and
pressure permitted determination of the density with uncertainty of $0.2\%$. 
The target
cell was surrounded by $24$ ``start counter'' scintillators that were used
in the event trigger. 

The CLAS detector utilized a non-uniform toroidal magnetic field of peak 
strength near $1.8$~T in conjunction with drift chamber tracking to determine 
particle momenta. The detector was divided into 6 sectors, such that when 
viewed along the beam line it was six-fold symmetric.
Charged particles with laboratory polar angles in the range 
$8^{\circ}-140^{\circ}$ could be tracked over approximately $83$\% of the
azimuthal angle.
A set of $288$ scintillators placed outside of the magnetic 
field region were used in the event trigger and during offline analysis in 
order to determine time of flight (TOF).
The momentum resolution of the detector was, on average, about $0.5$\%.
Other components of the CLAS, such as the Cerenkov counters and the 
electromagnetic calorimeters, were not used in this analysis.
A more detailed description of the CLAS can be found in Ref.~\cite{clas-detector}.

The event trigger required a coincidence between signals from the photon tagger
and the CLAS. 
The signal from the tagger consisted of an OR of the first $40$ of the $61$ total timing 
scintillators, corresponding to photon energies above $1.5$~GeV. Recording of
events associated with photons hitting counters $41$--$61$ required a random
tagger hit in counters $1$--$40$ during the trigger timing window. This allowed for
the acquisition of greater statistics at the higher photon-energy range of this
experiment.
The signal from
the CLAS required at least two sector-based signals. These signals 
consisted of an OR of any of the 4 start counter
scintillators in coincidence with an OR of any of the 48 time-of-flight
scintillators in the sector. The rate at which hadronic events were accumulated
was about 5~kHz; however, only a small fraction of these events contained the
reaction of interest to the analysis presented here.

\section{\label{section:data}Data and Event Selection}
The data used were obtained in the summer of 2004 during the CLAS ``g11a''
data taking period, in which approximately $20$ billion triggers were 
recorded. The relatively loose electronic trigger led to accumulation of
data for a number of photoproduction reactions.
The relative timing of the photon tagger, the start
counter and the time-of-flight elements were aligned during offline 
calibration. Calibrations were also made
for the drift times of each of the drift chamber packages and the pulse heights
of each of the time-of-flight counters. Finally, processing of the raw data
was performed in order to reconstruct tracks in the drift chambers and match 
them with time-of-flight counter hits. 

The reconstructed tracks were corrected for small imperfections in the magnetic 
field map and drift chamber alignment, along with  their mean energy losses as 
they passed through the target, the beam pipe, the start counter and air. In 
addition, small corrections were made to the incident photon energies to
account for mechanical sag in the tagger hodoscope.

The CLAS was optimized for detection of charged particles; thus, the 
$\pi^+\pi^-\pi^0$ decay of the $\omega$ was used to select the reaction of 
interest in this analysis. Detection of two positively charged particles and one 
negatively charged particle was required. A cut was placed on the confidence 
levels obtained from one-constraint kinematic fits performed to the hypothesis 
$\gamma p \rightarrow p \pi^+ \pi^- (\pi^0)$ in order to select events consistent 
with a missing $\pi^0$. All negatively charged tracks were assigned a $\pi^-$ 
identification. Kinematic fits were run for each of the possible $p,\pi^+$ 
particle assignments for the positively charged 
tracks using each of the recorded photons in the event. The combinations 
of photons and charged particles
with confidence levels greater than $10\%$ were retained for further analysis. 

The covariance matrix was studied using four-constraint kinematic fits
(energy and momentum conservation imposed) 
performed on the exclusive reaction $\gamma p \rightarrow p \pi^+ \pi^-$
in both real and Monte Carlo data samples. The confidence levels in all 
kinematic regions were found to be sufficiently flat and the pull-distributions 
(stretch functions) were Gaussians centered at zero with $\sigma = 1$
(see Fig.~\ref{fig:kfit}). The uncertainty in the extracted yields due to 
differences in signal lost because of this confidence-level cut in real as
compared to Monte Carlo data is estimated to be $3\%-4\%$.

The tagger signal time, which was synchronized with the accelerator radio-frequency 
(RF) timing, was propagated to the reaction vertex in order to obtain the start time for 
the event. The stop time for each track was obtained from the TOF scintillator
element hit by the track. The difference between these two times was the measured
time of flight, $t_{meas}$. Track reconstruction through the CLAS magnetic field yielded 
both the momentum, $\vec{p}$, of each track, along with the path length, $L$,
from the reaction vertex to the time-of-flight counter hit by the track. 
The expected time of flight for a mass hypothesis, $m$, is then given by
\begin{equation}
  t_{exp} = \frac{L}{c}\sqrt{1 + \left(\frac{m}{p}\right)^2}.
\end{equation}
The difference in these two time-of-flight calculations, 
$\Delta tof = t_{meas} - t_{exp}$, was used in order to separate protons from
pions and to remove events associated with out-of-time  photons.

\begin{figure}[h]
  \centering
  \includegraphics[width=0.49\textwidth]{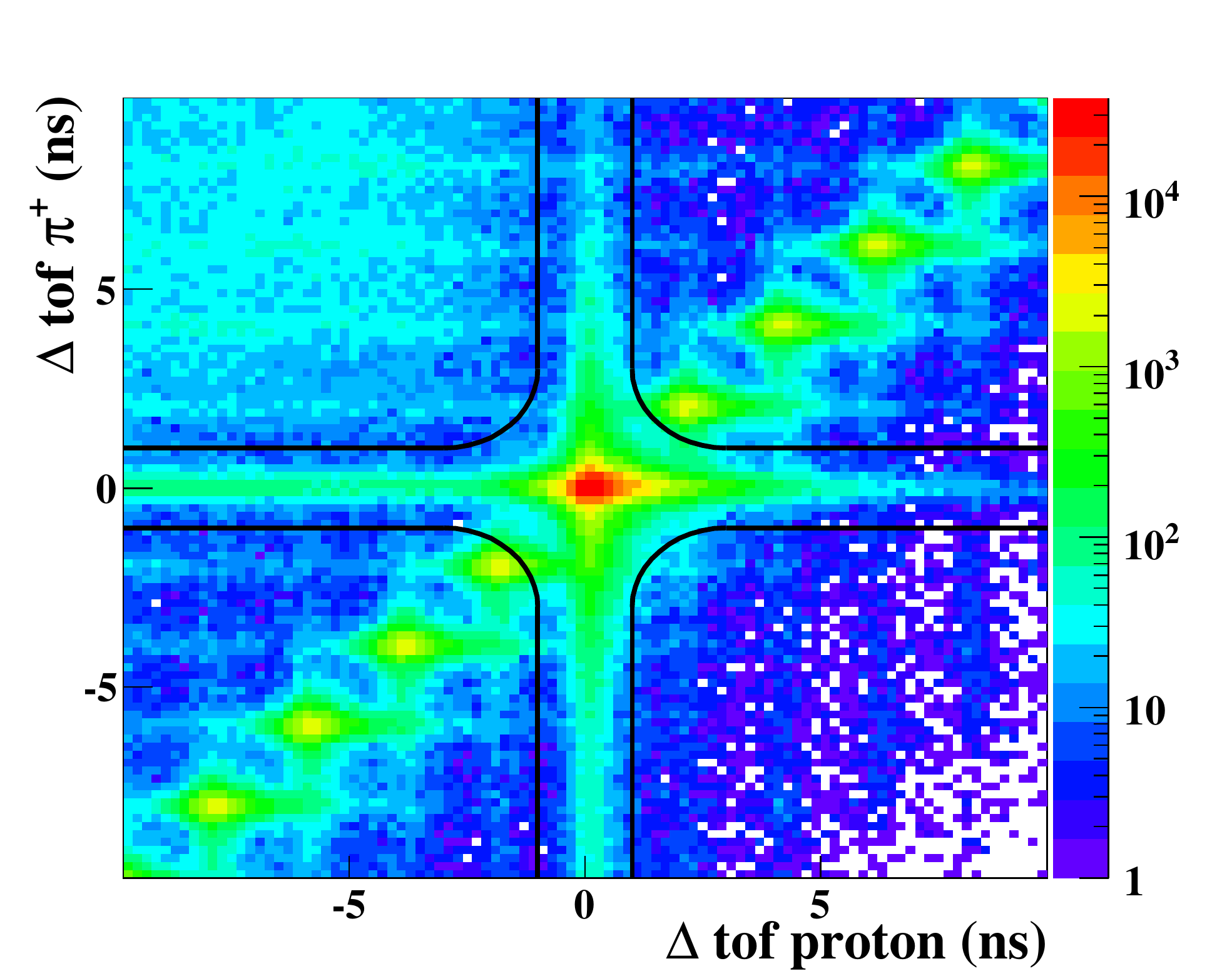}
\caption[]{\label{fig:pid}
  (Color Online)
  $\Delta tof_{\pi^+}$(ns) versus $\Delta tof_{p}$(ns):  
  Particle identification cut for a sample of events that pass a 10\% 
  confidence level cut when kinematically fit to the hypothesis
  $\gamma p \rightarrow p \pi^+ \pi^- (\pi^0)$.
  The black lines indicate the timing cuts. Note the logarithmic scale on the
  intensity axis.
}
\end{figure}

Fig.~\ref{fig:pid} shows $\Delta tof$ for the track passing the 
kinematic fit under the $\pi^+$ hypothesis versus $\Delta tof$ for the track passing the
fit under the proton hypothesis. The region near $(0,0)$ contains events where
both tracks are good matches to their respective particle identification 
hypotheses. The 2-ns radio-frequency 
time structure of the accelerator is evident in the out-of-time event clusters.
Events outside of the black lines, where neither hypothesis was met, were cut
from our analysis. This cut was designed to remove a minimal amount of good
events. The Feldman-Cousins
method~\cite{feldman} was used to place an upper limit on the signal 
lost 
at $1.3\%$. 
Any remaining accidental events 
fell into
the broad background under the $\omega$, and were rejected during the 
signal-background separation stage of the analysis discussed in 
Sec.~\ref{section:sig-bkgd}.

Fiducial cuts were applied on the momenta and angles of the tracks in order to
select events from
the well-understood regions of the detector. Included in these cuts was the
removal of $13$ of the $288$ time-of-flight elements due to poor performance.
In addition, events where the missing $\pi^0$ was moving along the beam line,
$\cos{\theta_{c.m.}^{\pi^0}} > 0.99$, 
were cut in order to remove leakage from the 
$\gamma p \rightarrow p \pi^+ \pi^-$ reaction.
A more detailed description of the analysis procedures presented in this article 
can be found in Ref.~\cite{williams-thesis}.

\begin{figure}
  \centering
\subfigure[]{
  \includegraphics[width=0.48\textwidth]{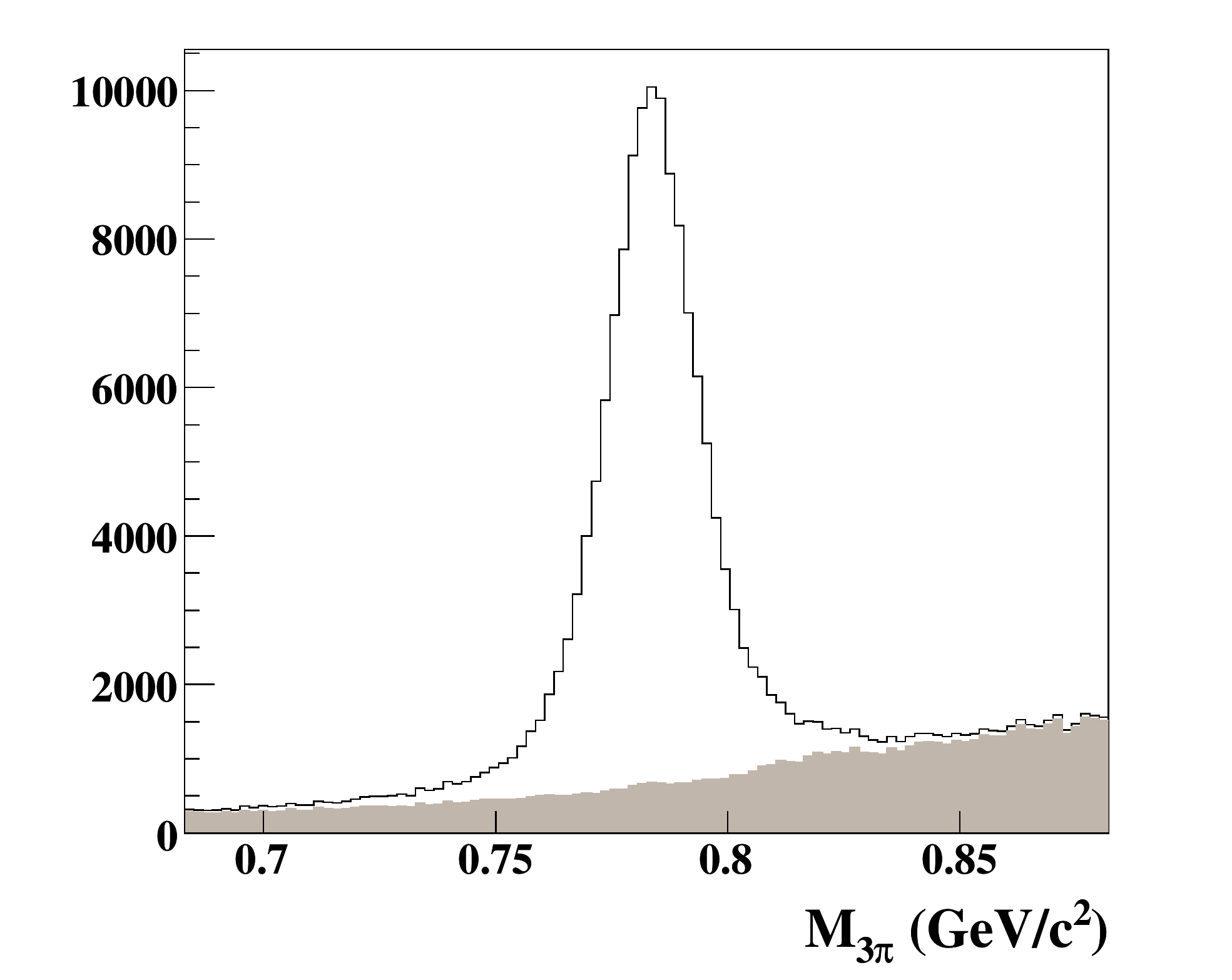}
} \\
\subfigure[]{
  \includegraphics[width=0.48\textwidth]{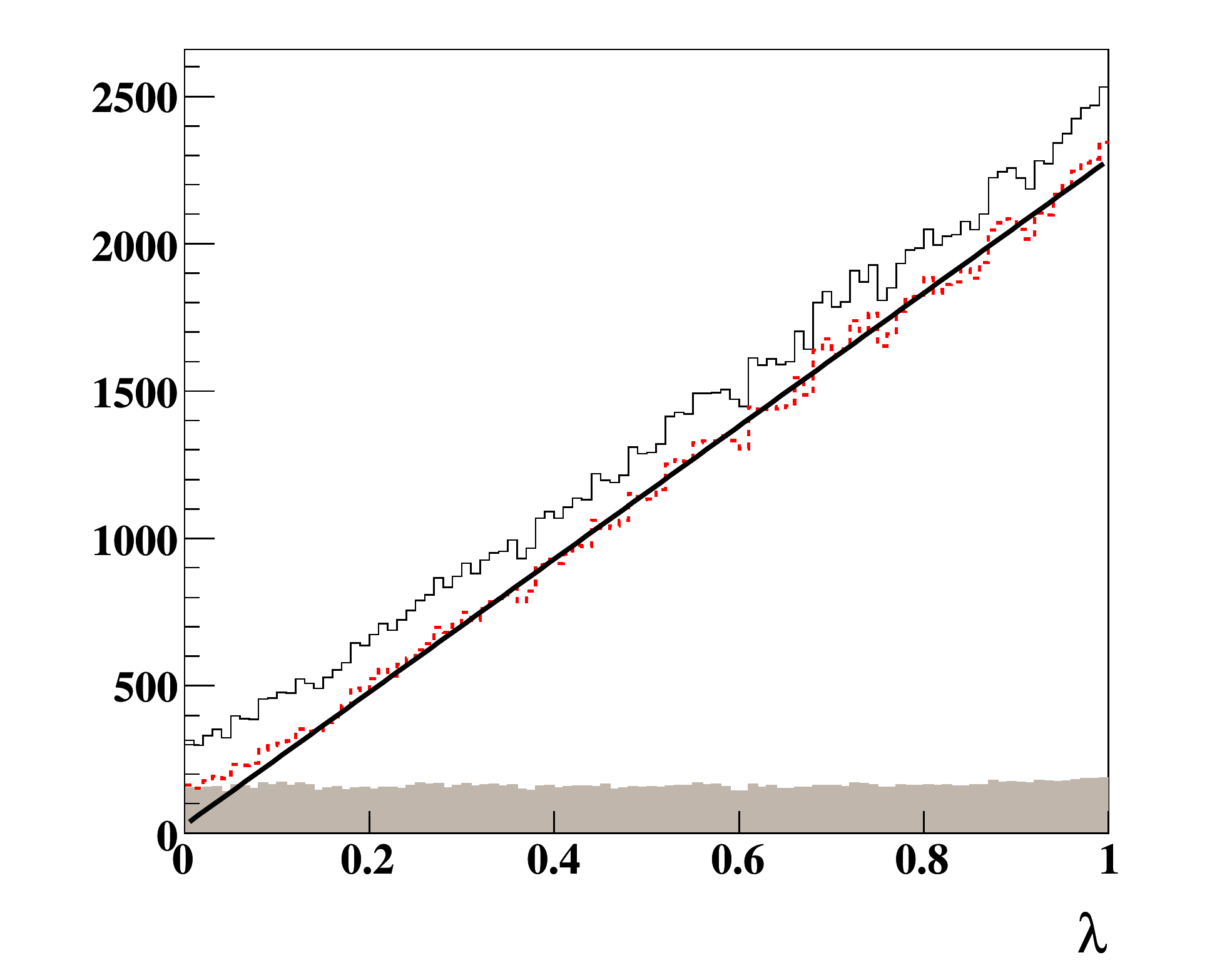}
}
\caption[]{\label{fig:sig-bkgd}
  (Color Online)
  (a) The $3\pi$ invariant mass distribution in the $W = 2.205$~GeV bin,
  integrated over all kinematics, for all events (unshaded) and for events 
  weighted by the background factors, $1-Q$ (shaded).
  (b) The $\lambda$ distribution of events in the same $W$ bin
  that satisfy $|M_{3\pi} - M_{\omega}| < 25$~MeV/$c^2$ (unshaded), the 
  same events weighted by signal factors $Q$ (dashed-red), and by 
  background factors $1-Q$ (shaded). The line represents a fit of the signal 
  to the function $a\lambda$.  
}
\end{figure}

\section{\label{section:sig-bkgd}Signal-Background Separation}
In addition to $\omega$ events, the resulting sample consisted 
of events from the reaction $\gamma p\rightarrow p\pi^{+}\pi^{-}\pi^{0}$ where 
the three-pion invariant mass was consistent with that of the $\omega$. 
These background events could arise from a variety of physics reactions, but 
they all share the characteristic that they cannot reproduce the narrow mass 
structure associated with the $\omega$. Typical background levels were $5$ to 
$10$\% of the $\omega$ peak height, while in a small number of bins near 
threshold and where the cross section is very low 
({\em e.g.} near $\cos\theta^{\omega}_{c.m.}$ of $-0.2$ at the highest photon
energies), the background was as large as $25$\% of the $\omega$ peak.
Thus, the three-pion invariant mass could be used as a tool to separate the 
signal from the background.

In order to accurately extract physical observables for $\omega$ 
photoproduction,
background events
(all non-$\omega$ events) must be separated from the signal in a way that
preserves all kinematic correlations. The method we have developed,
described in detail elsewhere~\cite{williams-thesis,bkgd-preprint}, was used
to assign each event a signal weight factor, $Q$, or equivalently, a 
background weight factor, $1-Q$. These $Q$-factors were then used to weight 
each
event's contribution to the ``log likelihood'' during the event-based fits 
discussed in Sec.~\ref{section:acc}. These fits were used to obtain the 
detector acceptance and to extract the spin density matrix elements. 
The $Q$-factors were also used to weight each event's contribution to the 
differential cross section.

The key feature of our procedure involves selecting each event's $N_c$ 
``nearest neighbor'' events (we chose $N_c = 500$). This is done by 
defining a metric in the space of all relevant kinematic variables, 
excluding the $3\pi$ invariant mass, $M_{3\pi}$.  
Each subset of events occupies a very small region of phase space; thus, the
$M_{3\pi}$ distribution can safely be used to determine each event's 
$Q$-factor,
while preserving the correlations present in the remaining kinematic variables.
This method facilitates separation of the signal and background without having 
to resort to dividing the data up into bins. 
Binning the data is undesirable due to the
high dimensionality of the reaction being studied in this analysis.

To this end, unbinned maximum likelihood fits were carried out for each event,
using its nearest neighbors, to determine the parameters 
$\vec{\alpha}= (b_0,b_1,b_2,b_3,b_4,s,\sigma)$ in the probability density
function
\begin{equation}
  F(M_{3\pi},\vec{\alpha}) 
  = \frac{B(M_{3\pi},\vec{\alpha}) + S(M_{3\pi},\vec{\alpha})}
  {\int \left(B(M_{3\pi},\vec{\alpha}) + S(M_{3\pi},\vec{\alpha})\right)
    dM_{3\pi}},
\end{equation}
where
\begin{equation}
  S(M_{3\pi},\vec{\alpha}) 
  = s\cdot V(M_{3\pi},M_{\omega},\Gamma_{\omega},\sigma)
\end{equation}
parametrizes the signal as a Voigtian (convolution of a 
Breit-Wigner and a Gaussian) with mass ${M_{\omega}=0.78256}$~GeV/$c^2$, 
natural width ${\Gamma_{\omega}=0.00844}$~GeV/$c^2$ and resolution
$\sigma$. The parameter $s$ sets the overall strength of the signal.
The background in each small phase space region was parametrized as a 
fourth order polynomial,
\begin{equation}
  B(M_{3\pi},\vec{\alpha}) = b_4 M_{3\pi}^4 + b_3 M_{3\pi}^3 + b_2 M_{3\pi}^2
  + b_1 M_{3\pi} + b_0.
\end{equation}
The $Q$-factor for the event was then calculated as
\begin{equation}
  Q_i = \frac{S(M_{3\pi}^i,\hat{\alpha}_i)}
  {S(M_{3\pi}^i,\hat{\alpha}_i) + B(M_{3\pi}^i,\hat{\alpha}_i)},
\end{equation}
where $M_{3\pi}^i$ is the event's $3\pi$ invariant mass and $\hat{\alpha}_i$ 
are the
estimators for the parameters obtained from the $i^{th}$ event's fit. The
signal yield could then be obtained in any kinematic bin as
\begin{equation}
  \mathcal{Y}_{\omega} = \sum\limits_{i}^{N} Q_i,
\end{equation}
where $N$ is the number of events in the bin.

The full covariance matrix obtained from each fit was used to
obtain the uncertainty in $Q$, $\sigma_{Q}$. 
This varied depending on kinematics; however, it was typically about 3\%. 
The uncertainty of the extracted yield, in any kinematic bin, 
was obtained by adding the $Q$-factor uncertainties (assuming 100\% correlation)
to the statistical uncertainty of the yield:
\begin{equation}
  \sigma^2_{\mathcal{Y_{\omega}}} = \sum\limits_i^{N} Q_i^2 + 
  \left(\sum\limits_i^{N} \sigma_{Qi}\right)^2.
\end{equation}
Studies were performed using various background
parametrizations, including polynomials of different orders, all of which 
yielded results within the values obtained for $\sigma_{Q}$. Therefore, we 
conclude that no additional systematic uncertainty is required.

Fig.~\ref{fig:sig-bkgd} demonstrates the effectiveness of applying this 
procedure in a single center-of-mass energy bin. 
Fig.~\ref{fig:sig-bkgd}(a) shows the $M_{3\pi}$ distribution, integrated 
over all kinematics, and the estimated background using the procedure 
described above. 
The results are quite plausible; however,
$\omega$ photoproduction provides us with a more stringent test of this 
procedure. 

The distribution of the decay quantity $\lambda$, which can be written in
terms of the pion momenta in the $\omega$ rest frame as
\begin{equation}
  \lambda = \frac{|\vec{p}_{\pi^+} \times \vec{p}_{\pi^-}|^2}
  {MAX\left(|\vec{p}_{\pi^+} \times \vec{p}_{\pi^-}|^2\right)},
\end{equation}
must be linear in shape and intersect 0 at $\lambda = 0$ for $\omega$ events 
--- this follows directly from the $\omega \rightarrow \pi^+ \pi^- \pi^0$ 
amplitude defined in Eq.~(\ref{eq:omega-decay-amp}).
Fig.~\ref{fig:sig-bkgd}(b) shows the $\lambda$ distribution, integrated 
over all kinematics, for events in the same bin shown in 
Fig.~\ref{fig:sig-bkgd}(a) in the region $\pm 25$~MeV/$c^2$ around the 
$\omega$ peak, 
along with the extracted signal and background distributions.
The signal distribution is well described by the function $a\lambda$. 
The small discrepancy near $\lambda = 0$ is the result of detector resolution. 

The method we have employed has cleanly separated signal from background
in the quantity $\lambda$, even though the known linear behavior of the signal
was not enforced in the fits. In fact, this method has effectively separated 
signal from background in all distributions, successfully preserving all 
kinematic correlations.

A detailed study of the systematic biases of the background subtraction
technique was carried out as part of this analysis. Not only was the function 
that was used to parametrize the background varied, but the number of nearest 
neighbor events was varied over a wide range and several different metrics were
used to determine the nearest neighbors events. The observed physical 
measurements were found to be completely insensitive to changes in these 
parameters over any reasonable set of values. Because of this, we associate 
no additional systematic error with these choices. 
A detailed description of this study is contained in Ref.~\cite{bkgd-preprint}.

\begin{figure}
  \centering
  \includegraphics[width=0.49\textwidth]{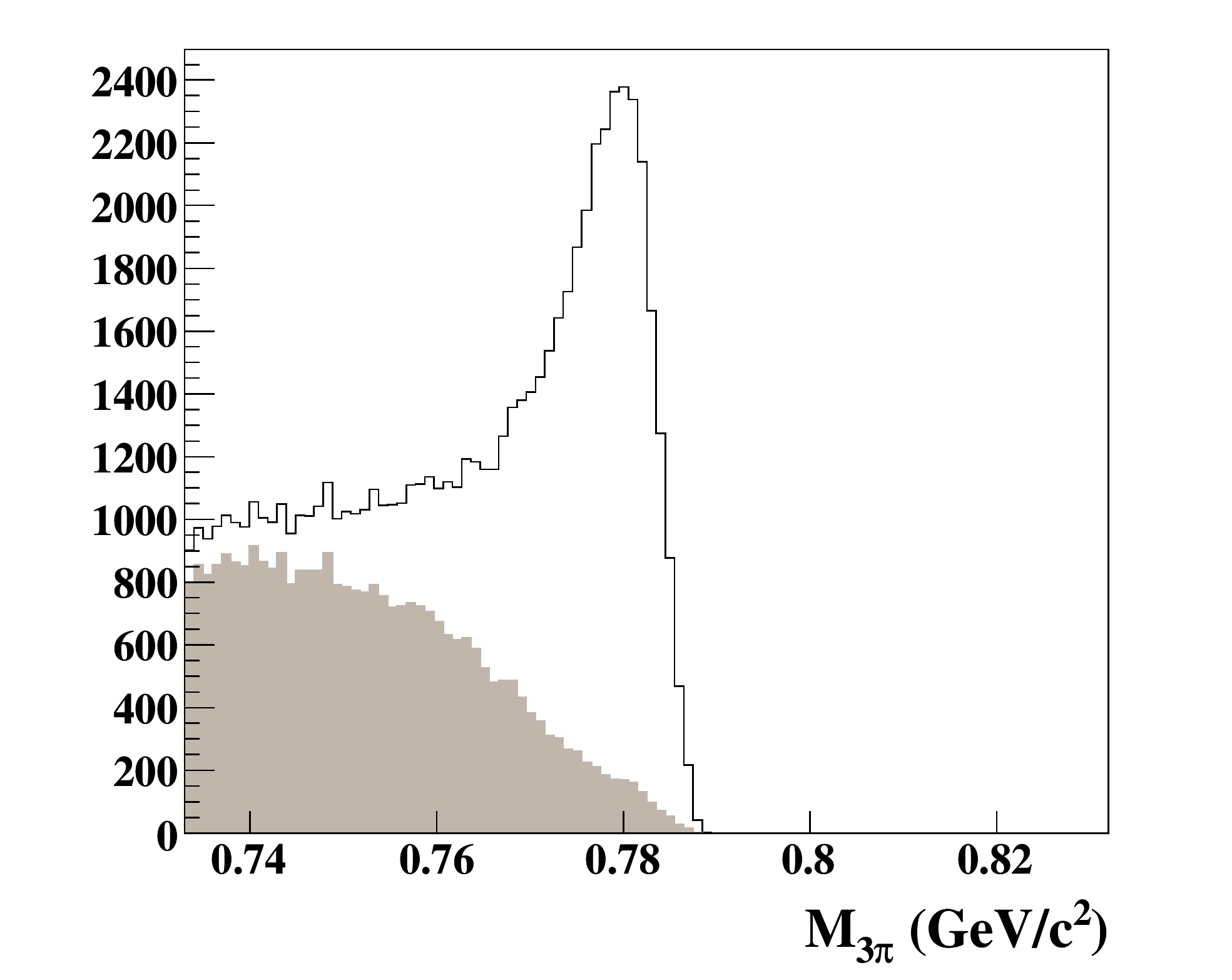}
\caption[]{\label{fig:sig-bkgd-threshold}
  The $3\pi$ invariant mass distribution in the ${W = 1.735}$~GeV bin, 
  integrated over all kinematics, for all events (unshaded) and for events 
  weighted by the background factors, $1-Q$ (shaded).  
}
\end{figure}

For energy bins near threshold and for ``edge'' regions ({\em i.e.} forward- 
and
backward-most angles) in some energy bins, the lack of events on both sides
of the peak leaves the fits unconstrained.
In these regions, the energy dependence of the 
$Q$-factors obtained in the closest energy bins for which fitting could be used
were projected down to the regions in question in order to obtain the 
$Q$-factors.
Fig.~\ref{fig:sig-bkgd-threshold} shows the results of this procedure in
the $W = 1.735$~GeV bin. By studying the $\lambda$ distributions in
these bins, the systematic uncertainty associated with the projected 
$Q$-factors
in the edge regions is estimated to be $5$\%. In the first two energy bins
above threshold, the uncertainties are estimated to be 15\% and 10\% for the
$W = 1.725$~GeV and $1.735$~GeV bins, respectively.

\section{\label{section:acc}Acceptance}

The efficiency of the detector was modeled using the standard CLAS GEANT-based 
simulation package and the Monte Carlo technique. A total of 200 million events
was generated pseudo-randomly, sampled from a phase space distribution. 
Each particle was propagated from the event vertex through the CLAS
resulting in a simulated set of detector signals for each track. The simulated
events were then processed using the same reconstruction software as the data.
In order to account for the event trigger used in this experiment 
(see Sec.~\ref{section:setup}), a study was performed to obtain the 
probability of a track satisfying the sector-based coincidences 
required by the trigger as a
function of kinematics and struck detector elements. The average effect of 
this correction in our analysis, which requires three detected particles, is
about 5\%--6\%.

An additional momentum smearing algorithm was applied in order to better match 
the resolution of the Monte Carlo to that of the data. Its effects were studied
using four-constraint kinematic fits performed on simulated 
$\gamma p \rightarrow p \pi^+ \pi^-$ events. After applying the momentum 
smearing algorithm, the same covariance matrix used for the data also
produced flat confidence level distributions in all kinematic regions for 
the Monte Carlo data as well.
The simulated $\omega$ events were then processed with the same analysis 
software as the data, including the one-constraint kinematic fits. At this
stage, all detector and software efficiencies were accounted for. 

In order to evaluate the CLAS acceptance for the 
${\gamma p \rightarrow p\omega}$ reaction, 
all kinematic correlations between the final state
particles must be accurately reproduced by the simulated data. 
Typically, this is done by using a 
physics model when generating Monte Carlo events. Due to the lack of any
pre-existing precise polarization measurements in the kinematic regions that
contain most of our data,
this was not an option. Instead, we chose to expand the 
scattering amplitude, $\mathcal{M}$, in a very large basis of $s$-channel 
waves as follows:
\begin{eqnarray}
  \label{eq:scattering-amp}
  \mathcal{M}_{m_i,m_{\gamma},m_f,m_{\omega}}(\vec{x},\vec{\alpha})
    \hspace{0.25\textwidth}\nonumber \\ \approx
    \sum\limits_{J = \frac{1}{2}}^{\frac{21}{2}}\sum\limits_{P = \pm}
    \mathcal{A}_{m_i,m_{\gamma},m_f,m_{\omega}}^{J^P}(\vec{x},\vec{\alpha}),
\end{eqnarray}
where $\vec{\alpha}$ denotes a vector of 108 fit parameters, 
$\vec{x}$ denotes the complete set of kinematic variables describing the 
reaction,
$m_i,m_{\gamma},m_f,m_{\omega}$ are the spin projections on the
incident photon direction in the center-of-mass frame,
and $\mathcal{A}$ are the $s$-channel 
partial wave amplitudes. 

The $\omega \rightarrow \pi^+ \pi^- \pi^0$ amplitude,
which is included in the $\mathcal{A}$'s above, can be written in terms of 
the isovectors and the 4-momenta of the pions, $\vec{I}_{\pi}$ and $p_{\pi}$
respectively, as well as the $\omega$ 4-momentum ($q$) and polarization
($\epsilon$) as
\begin{eqnarray}
  \label{eq:omega-decay-amp}
  A_{\omega \rightarrow \pi^+ \pi^- \pi^0} \propto 
  \left((\vec{I}_{\pi^+} \times \vec{I}_{\pi^0}) \cdot \vec{I}_{\pi^-} \right)
  \hspace{0.1\textwidth}\nonumber \\
  \hspace{-0.1in}
  \times \epsilon_{\mu\nu\alpha\beta}p_{\pi^+}^{\nu}p_{\pi^-}^{\alpha}
  p_{\pi^0}^{\beta} \epsilon^{\mu}(q,m_{\omega}),
\end{eqnarray}
which is fully symmetric under interchange of the three pions. For this reaction, where 
all final states contain $\omega\rightarrow\pi^{+}\pi^{-}\pi^{0}$, the isovector triple 
product simply contributes a factor to the global phase of all amplitudes.
In the $\omega$ rest frame, Eq.~(\ref{eq:omega-decay-amp}) simplifies to
\begin{equation}
    A_{\omega \rightarrow \pi^+ \pi^- \pi^0} \propto
    \left( \vec{p}_{\pi^+} \times \vec{p}_{\pi^-} \right)
    \cdot  \vec{\epsilon}(m_{\omega}).
\end{equation}
The remaining $s$-channel structure of the amplitudes $\mathcal{A}$, 
as well as the details
concerning the fit parameters, is described in~\cite{williams-thesis}.

\begin{figure}
  \centering
  \includegraphics[width=0.49\textwidth]{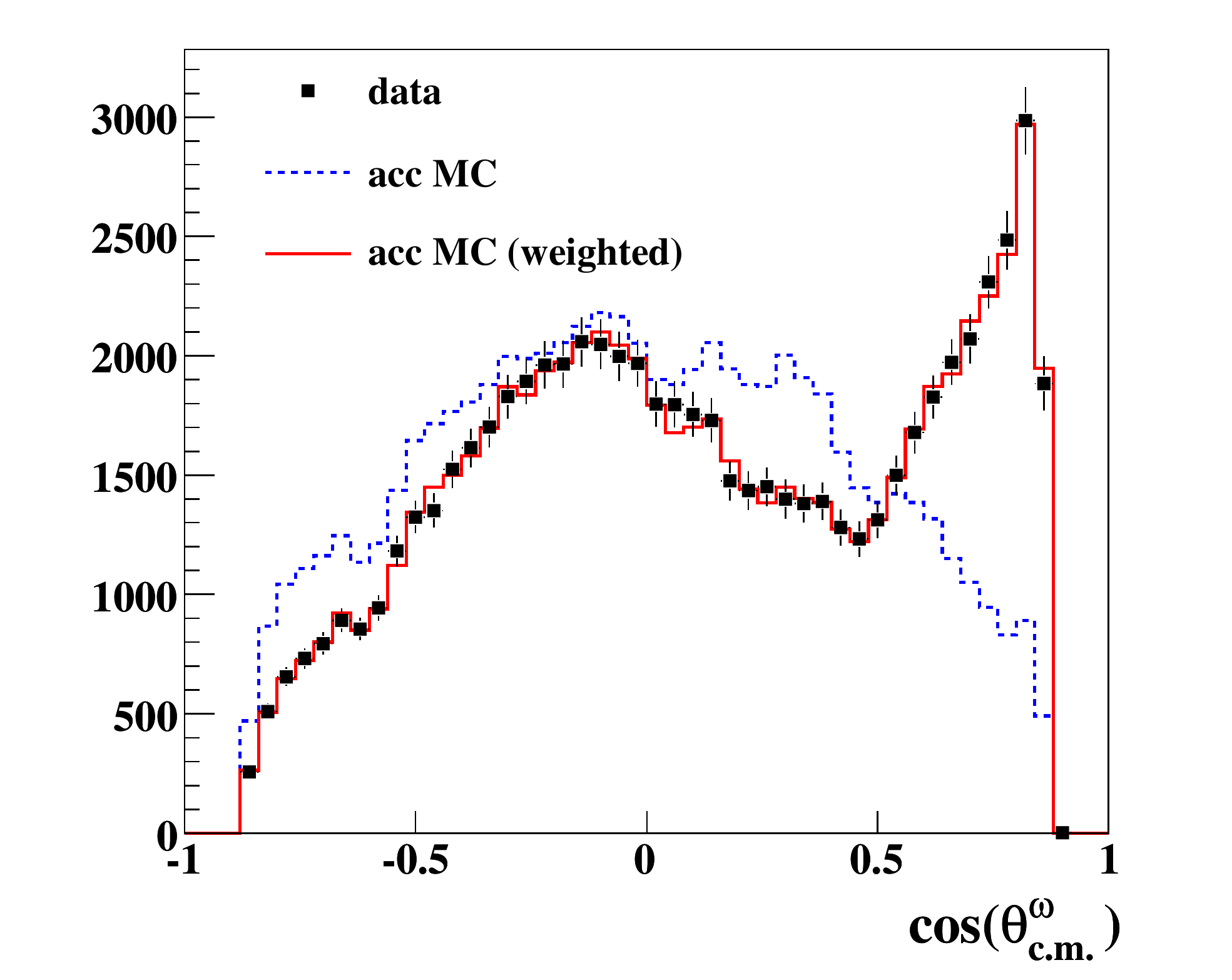}
\caption[]{\label{fig:wtd-acc-1d}
  (Color Online)
  Example fit result in the ${W = 2.005}$~GeV bin for 
  data (black squares), phase space accepted Monte Carlo events 
  (blue dashed line) and 
  phase space accepted Monte Carlo events weighted by the fits 
  discussed in Sec.~\ref{section:acc}
  (red solid line). 
  The weighted Monte Carlo provides an excellent description of the data.
}
\end{figure}

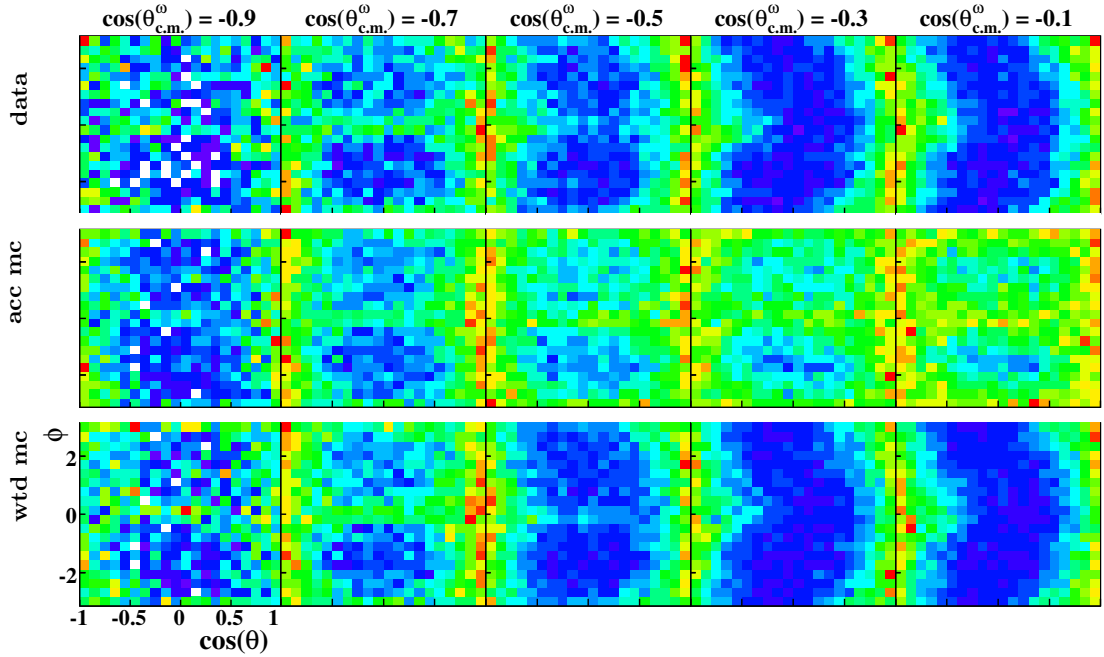
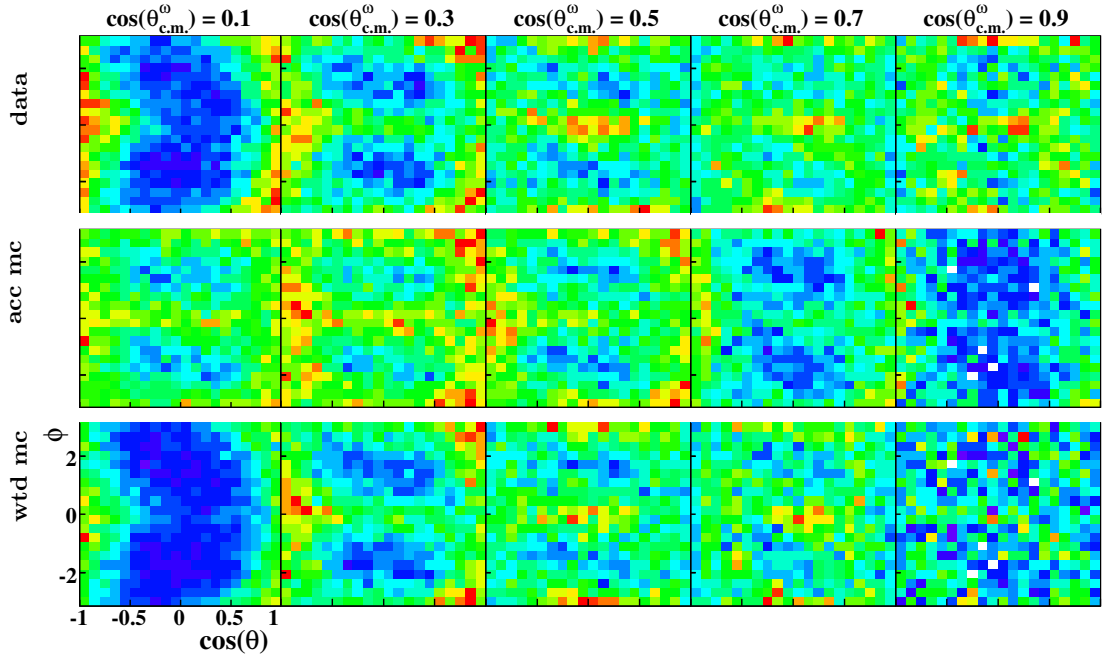
\begin{figure*}
  \centering
  \subfigure[]{\label{fig:wtd-acc-a}
    \includegraphics[width=0.80\textwidth]{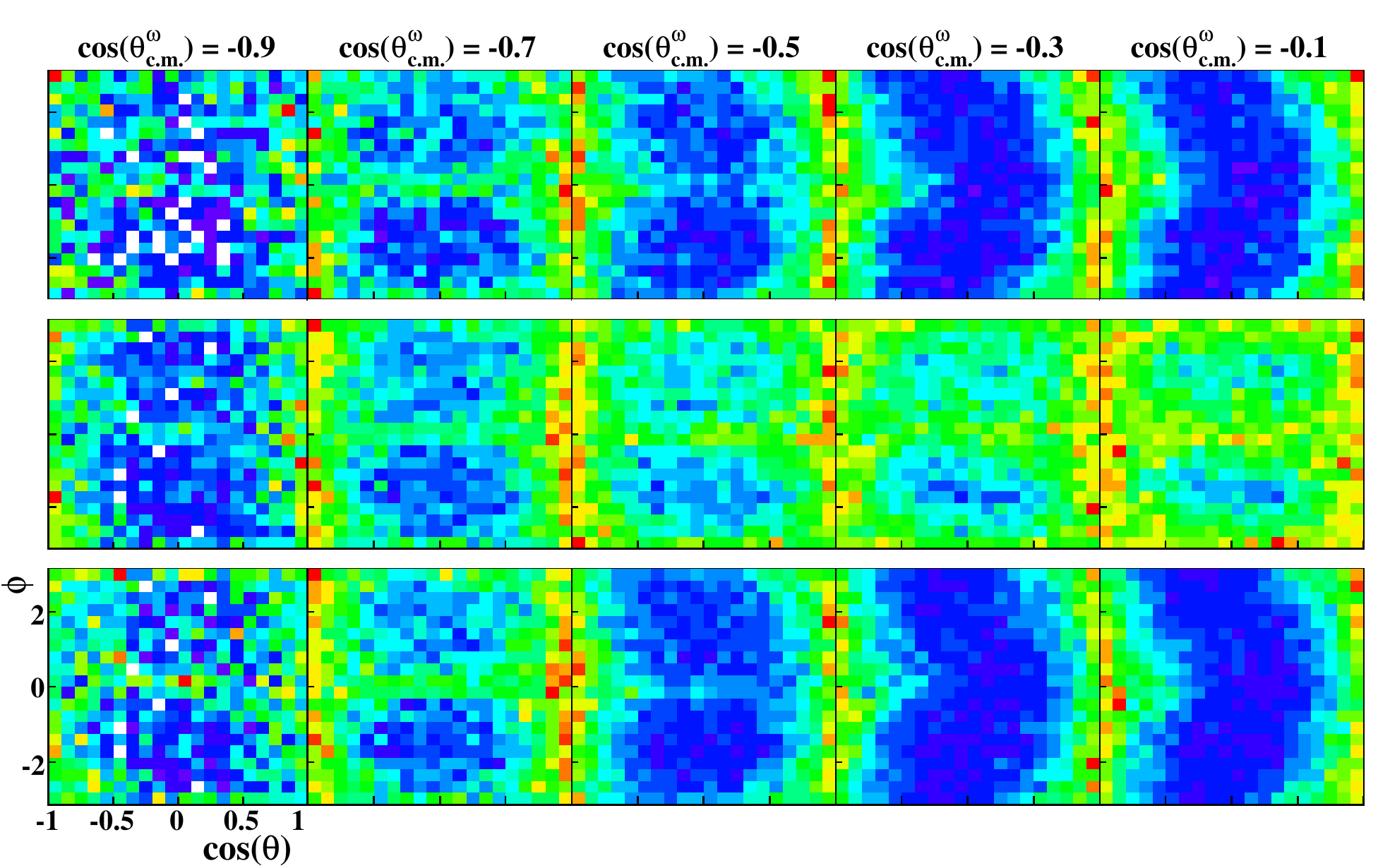}
    \put(-420,200) {\begin{sideways}\textbf{data}\end{sideways}}
    \put(-420,125) {\begin{sideways}\textbf{acc mc}\end{sideways}}
   \put(-420,50) {\begin{sideways}\textbf{wtd mc}\end{sideways}}
 }
  \subfigure[]{\label{fig:wtd-acc-b}
    \includegraphics[width=0.80\textwidth]{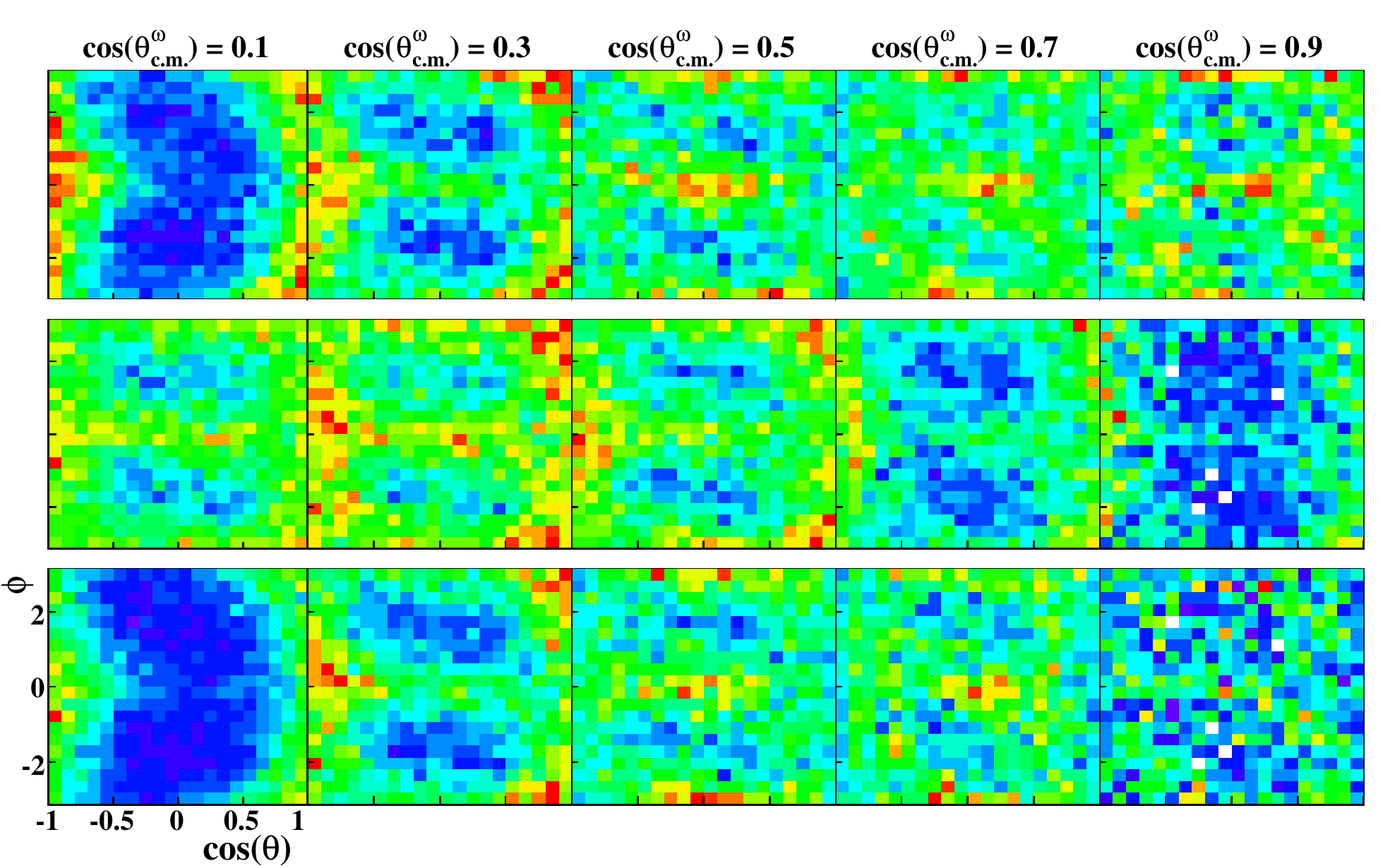}
    \put(-420,200) {\begin{sideways}\textbf{data}\end{sideways}}
    \put(-420,125) {\begin{sideways}\textbf{acc mc}\end{sideways}}
   \put(-420,50) {\begin{sideways}\textbf{wtd mc}\end{sideways}}
   }
\caption[]{\label{fig:wtd-acc}
  (Color Online)  $\phi$ versus $\cos{\theta}$ in the Adair frame (see Sec.~\ref{section:results}~B) 
  in $\cos{\theta^{\omega}_{c.m.}}$ bins: Example fit result in the $W = 2.005$~GeV bin for 
  data (top row), phase space accepted Monte Carlo events (middle row) and phase space accepted 
  Monte Carlo events weighted by the fits discussed in Sec.~\ref{section:acc} (bottom row). Panel (a)
  shows backwards angles, while panel (b) shows forward angles. Note that the weighted Monte Carlo 
  provides an accurate description of the data ($\chi^2/$degrees-of-freedom~$\sim 0.6$).}
\end{figure*}

Unbinned maximum likelihood fits were performed in each $W$ bin in order
to obtain the estimators $\hat{\alpha}$ for the parameters $\vec{\alpha}$ in 
Eq.~(\ref{eq:scattering-amp}). The results of these fits were used to obtain a
weight, $I_i$, for each Monte Carlo event according to
\begin{equation}
  I_i = \sum\limits_{m_i,m_{\gamma},m_f}\left\vert \sum\limits_{m_{\omega}}
  \mathcal{M}_{m_i,m_{\gamma},m_f,m_{\omega}}(\vec{x}_i,\hat{\alpha}_i)
  \right\vert^2,
\end{equation}
where $\vec{x}_i$ is the complete set of kinematic variables of the $i^{th}$ 
event. The
weighted accepted Monte Carlo events fully reproduce the real data in 
all distributions, including all correlations 
(see Figs.~\ref{fig:wtd-acc-1d} and \ref{fig:wtd-acc}).
We note here that the results of these fits are not interpreted as physics,
{\em i.e.} they are not considered evidence of resonance contributions to
the $\omega$ photoproduction reaction. 
They are simply used in order to provide a complete 
description of the data.

The acceptance in any kinematic bin is then
obtained as
\begin{equation}
  {\rm acc}(\vec{x}) = \frac{\sum\limits_i^{N_{acc}} 
    I_i}{\sum\limits_j^{N_{gen}} I_j},
\end{equation}
where $N_{acc} (N_{gen})$ is the number of accepted (generated) Monte Carlo 
events in the bin and the $I$'s are the event weights discussed above.
An accurate physics generator would use the factors of $I$ during
the event generation stage, rather than weighting the accepted events.
The resulting acceptance calculation would be the same, modulo statistical 
fluctuations.

The statistical uncertainties in the acceptance calculations due to the finite
number of Monte Carlo events generated in each kinematic bin were obtained
from the standard binomial distribution.
The systematic uncertainty in the acceptance calculation is discussed in
Sec.~\ref{section:errors}.

\section{\label{section:norm}Normalization}

The measured rate of electrons detected by the tagger was used to compute
the number of photons incident on the target by sampling tagger hits not in
coincidence with the CLAS. These rates were integrated over the
live-time of the experiment in order to obtain the total photon flux associated
with
each tagger element. Losses of photons in the beam line due to effects such
as beam collimation were determined during dedicated runs using a 
total-absorption counter placed downstream of the 
CLAS~\cite{gflux}.

The standard electronics hardware that has traditionally been used to 
determine the experimental live-time 
malfunctioned during the ``g11a'' data taking period.
A downstream device used to measure the beam current during 
electron runs~\cite{clas-detector} was used instead. The relatively low count
rate in this device during photon running led to increased uncertainty in the
live-time measurement. The stability of normalized $\omega$ yields for runs
with different beam currents was used to estimate this uncertainty to be about 
3\%.

As was stated in Sec.~\ref{section:setup}, only 40 of the 61 timing elements
of the photon tagger were included in the event trigger. Events 
associated with hits in the  ``untriggered'' counters, 41-61, were only 
recorded if a random hit in 
counters 1-40 occurred during the trigger time window. The electron rates
used to measure the photon flux, discussed above, were used to calculate the
probability of such an occurrence, $P_{trig} = 0.467$. 
The measured flux for tagger counters 41--61 was scaled down by $P_{trig}$ to 
account for the event trigger.

Defective electronics in one of the tagger channels led to inaccurate flux 
measurements in the energy bins at ${W=2.735}$ and 2.745~GeV. The flux in the
energy bin at ${W=1.955}$~GeV was also deemed unreliable due to its inclusion 
of events
associated with both triggered and untriggered tagger counters. 
Differential cross sections are not reported in these three energy bins;
however, spin density matrix elements, which do not require normalization 
information, are reported.
\section{\label{section:errors}Systematic Uncertainties}

The $\omega$ photoproduction cross section, for the case with an unpolarized
beam and an unpolarized target, must be isotropic in the azimuthal angle. 
Thus, the 
acceptance-corrected $\omega$ yields, each obtained in an individual CLAS 
sector,  must be consistent with each other. 
By examining the consistency of these yields, we estimated the relative 
uncertainty in the acceptance correction to be between 4\%--6\%, depending on
center-of-mass energy.
This is added in quadrature with uncertainties due to particle identification 
(1.3\%) and 
confidence level (3\%) cuts to obtain an overall estimated acceptance 
uncertainty of 5\%--7\%.

It is common practice in photoproduction experiments to check the 
quality of
the normalization calculation by computing the single pion cross section and
comparing it to the world's data; however,
the two-track trigger used in this experiment does not permit such a
calculation. In order to check our normalization, cross sections were also 
computed for several other reactions from the ``g11a'' data set
and compared to previously published CLAS data. 
The run-to-run consistency of the normalized $\omega$ yield was also examined. 
Based on these studies, we estimate the normalization uncertainty to be 7.3\%.
When combined with contributions from photon transmission efficiency (0.5\%) 
and live-time (3\%), the total
estimated normalization uncertainty is 7.9\%.

The acceptance and normalization uncertainties discussed above were then
combined with contributions from
target density and length (0.2\%), along with 
branching fraction (0.7\%) to obtain a total uncertainty, excluding 
contributions from signal-background separation that are calculated 
``point-to-point'', of about 9\%--11\%. In the first two energy bins above 
threshold, the additional uncertainties in the signal-background separation
method (see Sec.~\ref{section:sig-bkgd}) increase this number to 
13\%--17\%.
\section{\label{section:results}Results}
\subsection{\label{section:results:dsigma}Differential cross sections}
\begin{figure*}[p]
  \centering
  \hspace{-0.045\textwidth}
  \includegraphics[width=0.99\textwidth]{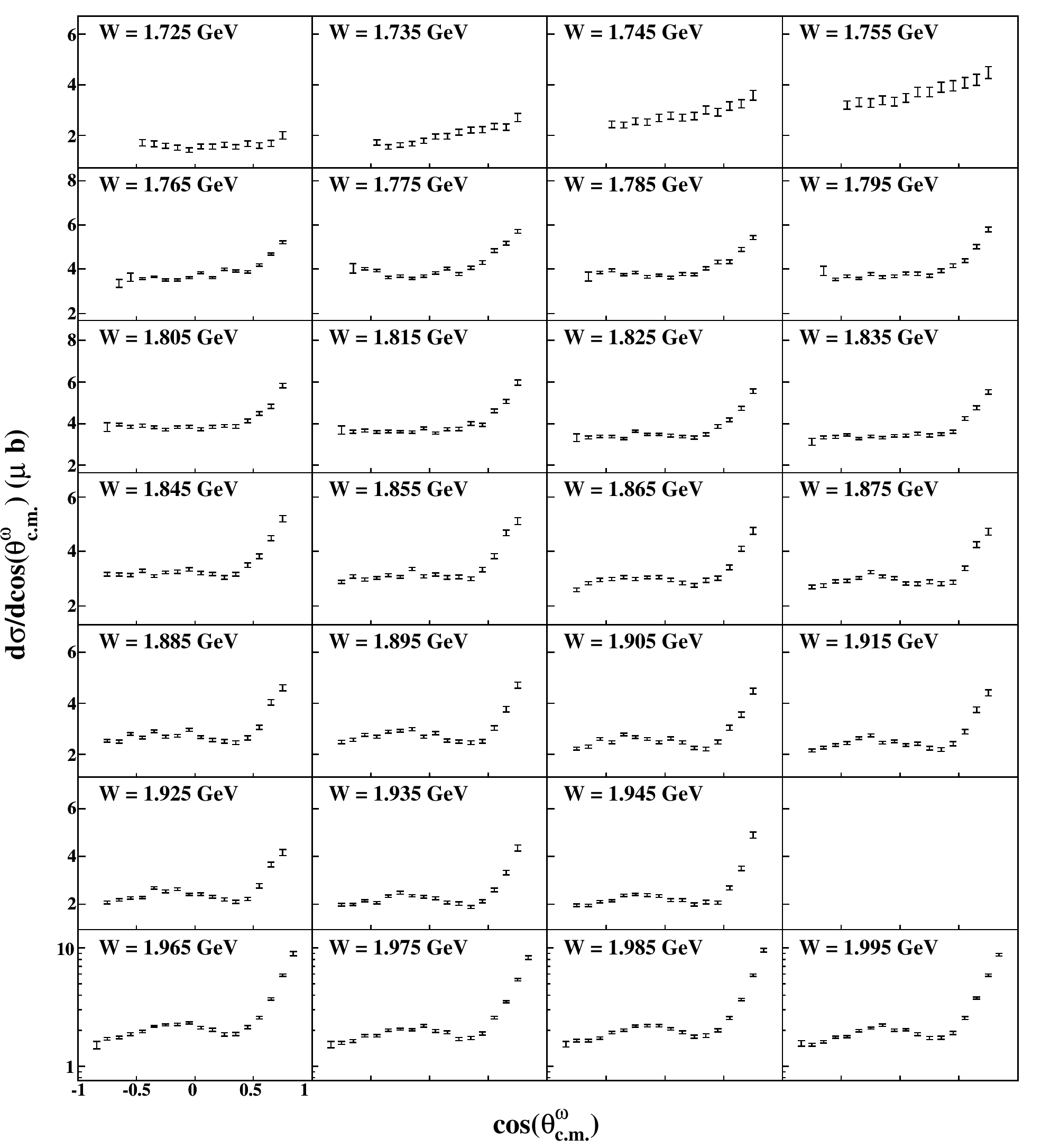}
\caption[]{\label{fig:dsigma-dcos-0}
  $\frac{d\sigma}{d\cos{\theta^{\omega}_{c.m.}}}$ $(\mu b)$ versus
  $\cos{\theta^{\omega}_{c.m.}}$: 
  Differential cross-section results for bins in the energy range  
  $1.72$~GeV~$\leq W <$~$2.00$~GeV. The centroid of each $10$-MeV wide bin is 
  printed on the plot. The lack of reported data points in the $W = 1.955$~GeV bin is 
  discussed in Sec.~\ref{section:norm}.  The error bars, which do not include systematic 
  uncertainties, are discussed in the text.  The additional near-threshold background 
  separation uncertainties, discussed in Sec.~\ref{section:sig-bkgd}, are clearly visible in the 
  first four center-of-mass energy bins. Note that the vertical scales are linear up to
 $W$ of $1.945$~GeV and logarithmic above that.}
\end{figure*}

\begin{figure*}[p]
  \centering
  \hspace{-0.045\textwidth}
  \includegraphics[width=0.99\textwidth]{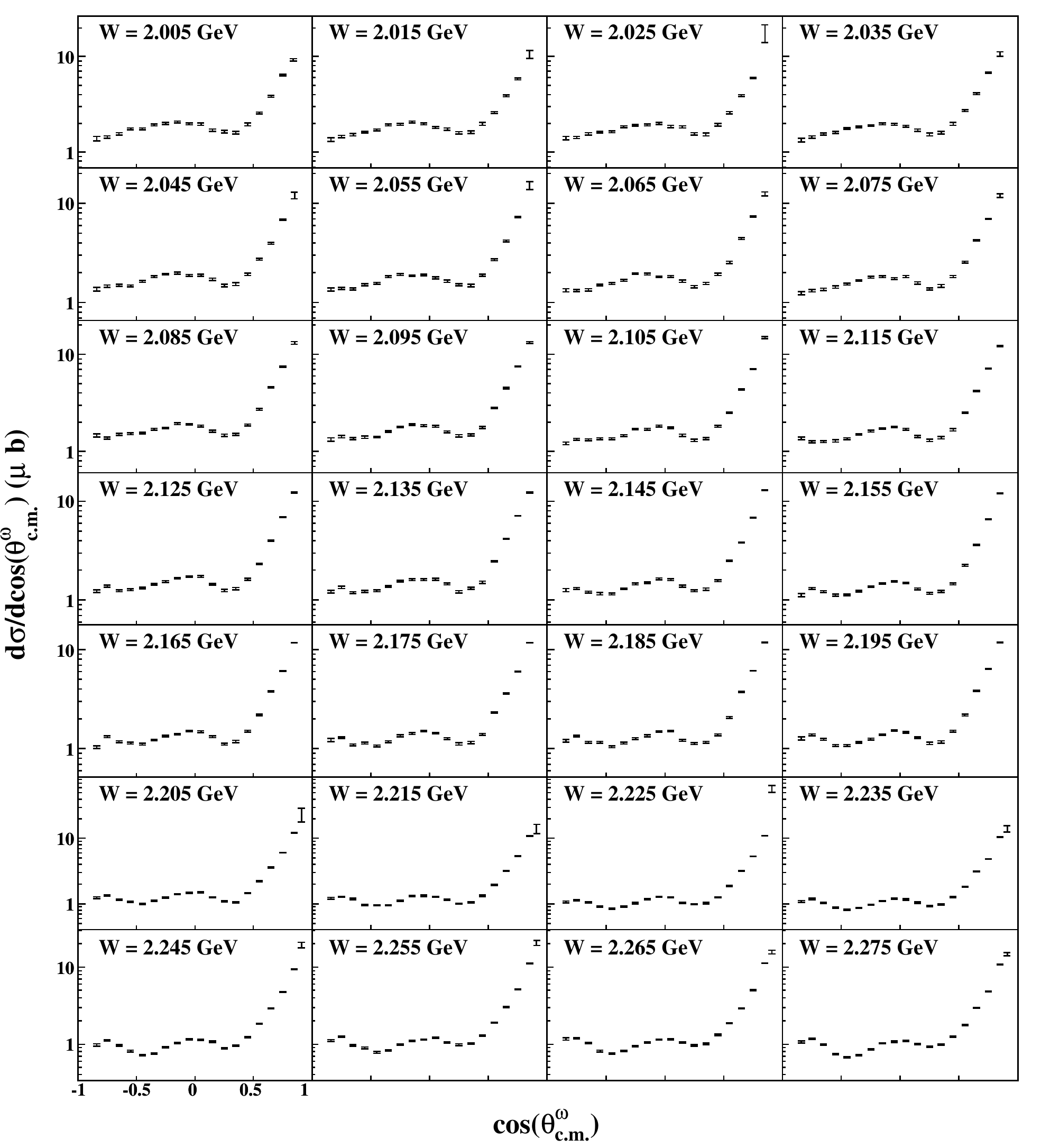}
\caption[]{\label{fig:dsigma-dcos-1}
  $\frac{d\sigma}{d\cos{\theta^{\omega}_{c.m.}}}$ $(\mu b)$ versus
  $\cos{\theta^{\omega}_{c.m.}}$: 
  Differential cross-section results for bins in the energy range
  $2.00$~GeV~$\leq W <$~$2.28$~GeV. The centroid of each $10$-MeV wide bin is 
  printed on the plot. The error bars, which do not include systematic uncertainties, are 
  discussed in the text.
}
\end{figure*}

\begin{figure*}[p]
  \centering
  \hspace{-0.045\textwidth}
  \includegraphics[width=0.99\textwidth]{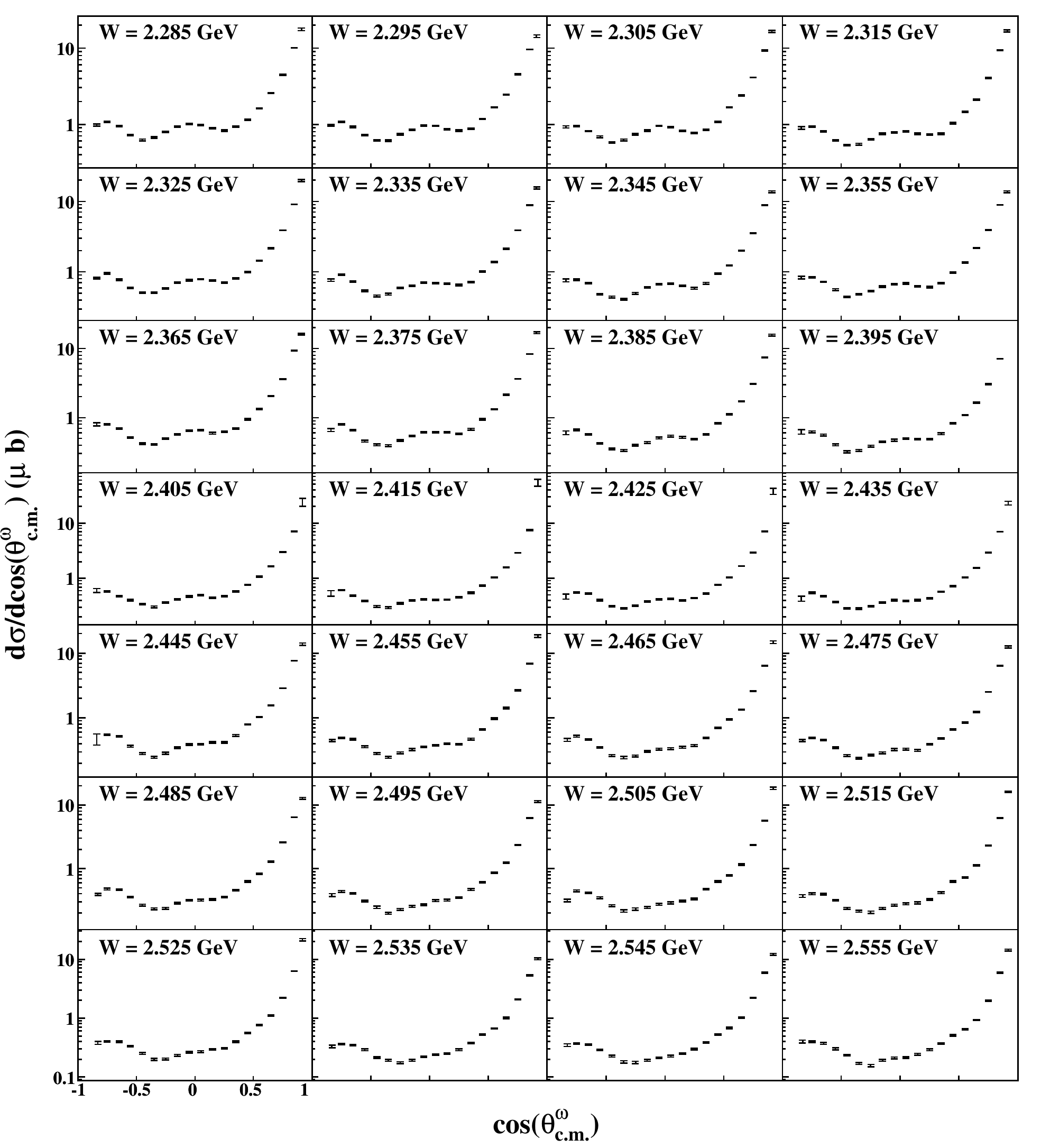}
\caption[]{\label{fig:dsigma-dcos-2}
  $\frac{d\sigma}{d\cos{\theta^{\omega}_{c.m.}}}$ $(\mu b)$ versus $\cos{\theta^{\omega}_{c.m.}}$: 
  Differential cross-section results for bins in the energy range
  $2.28$~GeV~$\leq W <$~$2.56$~GeV. The centroid of each $10$-MeV wide bin is 
  printed on the plot. The error bars, which do not include systematic uncertainties, are 
  discussed in the text.
}
\end{figure*}

\begin{figure*}[p]
  \centering
  \hspace{-0.045\textwidth}
  \includegraphics[width=0.99\textwidth]{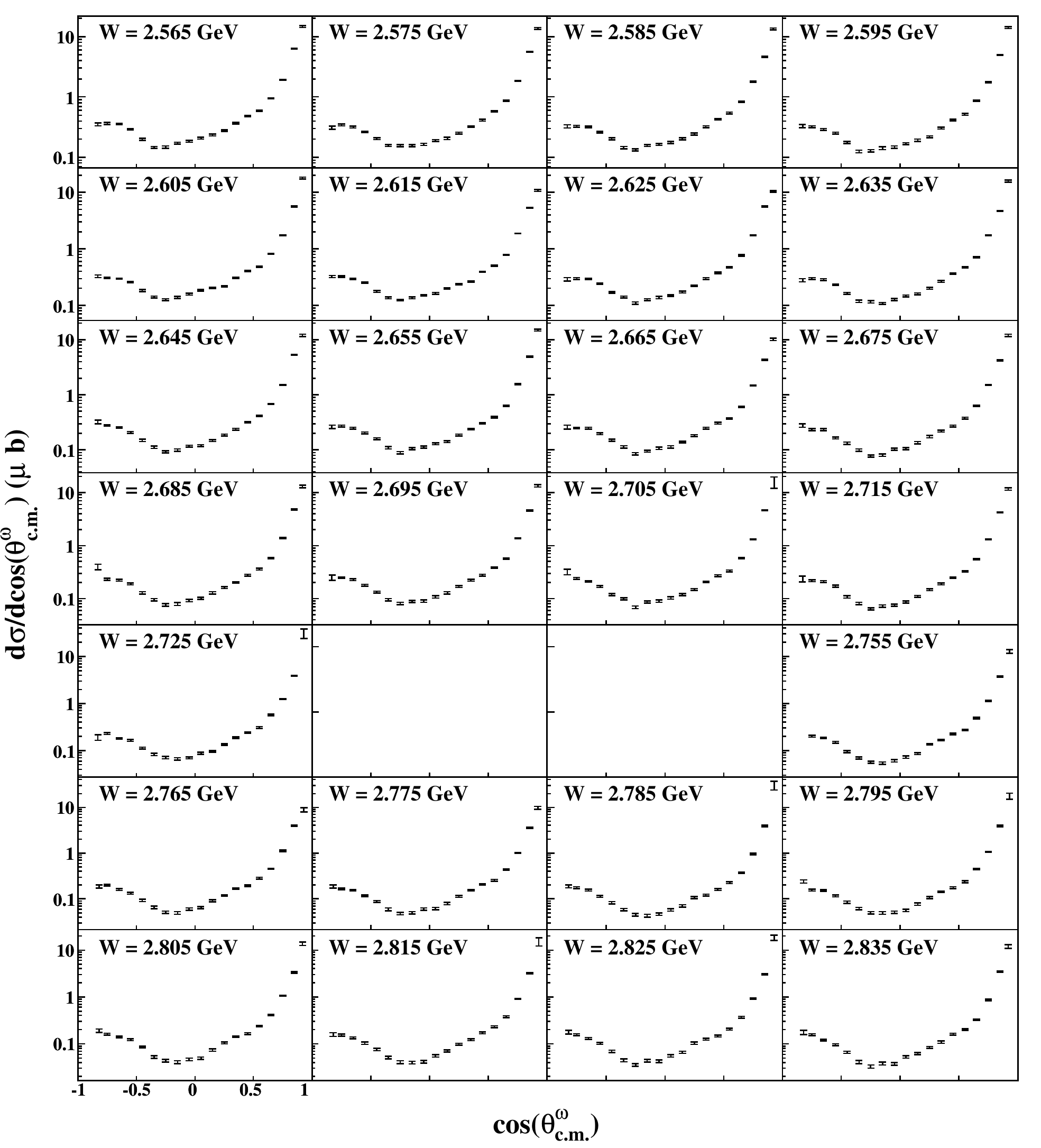}
\caption[]{\label{fig:dsigma-dcos-3}
  $\frac{d\sigma}{d\cos{\theta^{\omega}_{c.m.}}}$ $(\mu b)$ versus $\cos{\theta^{\omega}_{c.m.}}$: 
  Differential cross-section results for bins in the energy range 
  $2.56$~GeV~$\leq W <$~$2.84$~GeV. The centroid of each $10$-MeV wide bin is printed 
  on the plot. The lack of reported data points in the $W = 2.735$~GeV and $W = 2.745$~GeV 
  bins is discussed in Sec.~\ref{section:norm}. The error bars, which do not include systematic 
  uncertainties, are discussed in the text.
}
\end{figure*}

Differential cross sections, $d\sigma/d\cos{\theta^{\omega}_{c.m.}}$, were 
computed in 109 10-MeV wide bins in $W$. Each energy bin was divided 
into 20 bins in $\cos{\theta^{\omega}_{c.m.}}$ of width 0.1, although results 
could not be extracted in every bin due to limitations in the detector 
acceptance. In total, 1960 cross-section points are reported here.
The centroid of each bin is reported as the mean of the range of the bin
with nonzero acceptance. The results are shown in 
Figs.~\ref{fig:dsigma-dcos-0}, \ref{fig:dsigma-dcos-1}, \ref{fig:dsigma-dcos-2} and
\ref{fig:dsigma-dcos-3}. The error bars contain the uncertainties of the yield
extraction, discussed in Sec.~\ref{section:sig-bkgd}, along with statistical uncertainties from 
the Monte Carlo acceptance calculations. The overall systematic 
uncertainty, discussed in Sec.~\ref{section:errors}, is estimated to be
between 9\%--11\%, depending on center-of-mass energy.

In the ``transverse direction'', which we can loosely define as 
$|\cos{\theta^{\omega}_{c.m.}}| < 0.8$, there are several prominent features
present in the data. 
Near threshold, the transverse cross section is mostly flat. Around 
$W \sim 1.9$~GeV it begins to develop a {\em humped} shape and by 
$W \sim 2.1$~GeV the cross section has two dips. In a concurrent article, we 
present partial wave analysis results obtained from this data which attribute 
these features to various baryon resonance contributions~\cite{williams-prd}. 
For now, we simply aim to draw attention to some of the prominent structures 
present in our measurements.

A very prominent forward peak begins to rise just above threshold and continues
to be the dominant feature of the cross section up through our highest 
energies. This type of behavior typically indicates the presence of strong 
$t$-channel contributions. Models of $\omega$ photoproduction, 
{\em e.g.} Refs.~\cite{laget-2000,laget-2002,oh-2001,sibirtsev-2002}, 
typically associate this peak with $t$-channel contributions from $\pi^0$,
$\eta$ and Pomeron exchange.  
A backwards peak begins to emerge around $W \sim 2.2$~GeV, whose prominence 
increases as the energy increases (although it is always at least one order of 
magnitude smaller than the forward peak). This could be indicative of the 
presence of contributions in the $u$-channel.  Many models of this reaction
attribute this peak to $u$-channel nucleon 
exchange~\cite{laget-2000,laget-2002,oh-2001,sibirtsev-2002}; however, 
comparisons of the spin density matrix elements predicted by these models to
the new high-precision measurements presented in this article casts doubt on
the validity of these models (see Sec.~\ref{sec:interpretation}).

\subsection{Spin density matrix elements}

The polarization of the $\omega$ can be studied by examining the distributions
of its decay products. Since the $\omega$ is a spin-1 particle, its spin 
density matrix has nine complex elements; however, parity, hermiticity and 
normalization 
reduce the number of independent elements (for an unpolarized beam)
to four real quantities (of which, three are measurable). 
Traditionally, these are chosen
to be $\rho^0_{00}$, $\rho^0_{1-1}$ and $Re(\rho^0_{10})$. Our results
cover a large range of energies and angles; thus, we chose the quantization 
axis to
be the photon direction in the overall c.m. frame, known as the {\em Adair}
frame~\cite{schilling-1970}.

The spin density matrix elements can be written in terms of the production
amplitudes $\mathcal{A}$
({\em i.e.} the scattering amplitudes $\mathcal{M}$ introduced in
Sec.~\ref{section:acc} without the $\omega$ decay piece), 
as
\begin{equation}
  \rho^0_{MM'} = \frac{1}{N}\sum\limits_{m_{\gamma},m_i,m_f} 
  \mathcal{A}_{m_i,m_{\gamma},m_f,M} 
  \mathcal{A}^{*}_{m_i,m_{\gamma},m_f,M'},
\end{equation}
where the $M,M'$ refer to the spin projection of the $\omega$ 
(on the photon direction in the c.m. frame) and
\begin{equation}
  N = \sum\limits_{m_i,m_{\gamma},m_f} \sum\limits_{M} 
  |\mathcal{A}_{m_i,m_{\gamma},m_f,M}|^2,
\end{equation}
is a normalization factor. Using the production amplitudes obtained from the
event-based fits described in Sec.~\ref{section:acc}, the spin density 
matrix elements were projected out of the partial wave expansion at 2015 
$(W,\cos{\theta^{\omega}_{c.m.}})$ points. These data points correspond to
the centroids of the bins for which cross-section results are reported,
along with additional points in the $W = 1.955$~GeV, 2.735~GeV and 
2.745~GeV center-of-mass energy bins for which cross-sections results are not
reported due to normalization issues (see Sec.~\ref{section:norm}). 

\begin{figure*}[p]
  \centering
 \hspace{-0.045\textwidth}
  \includegraphics[width=0.99\textwidth]{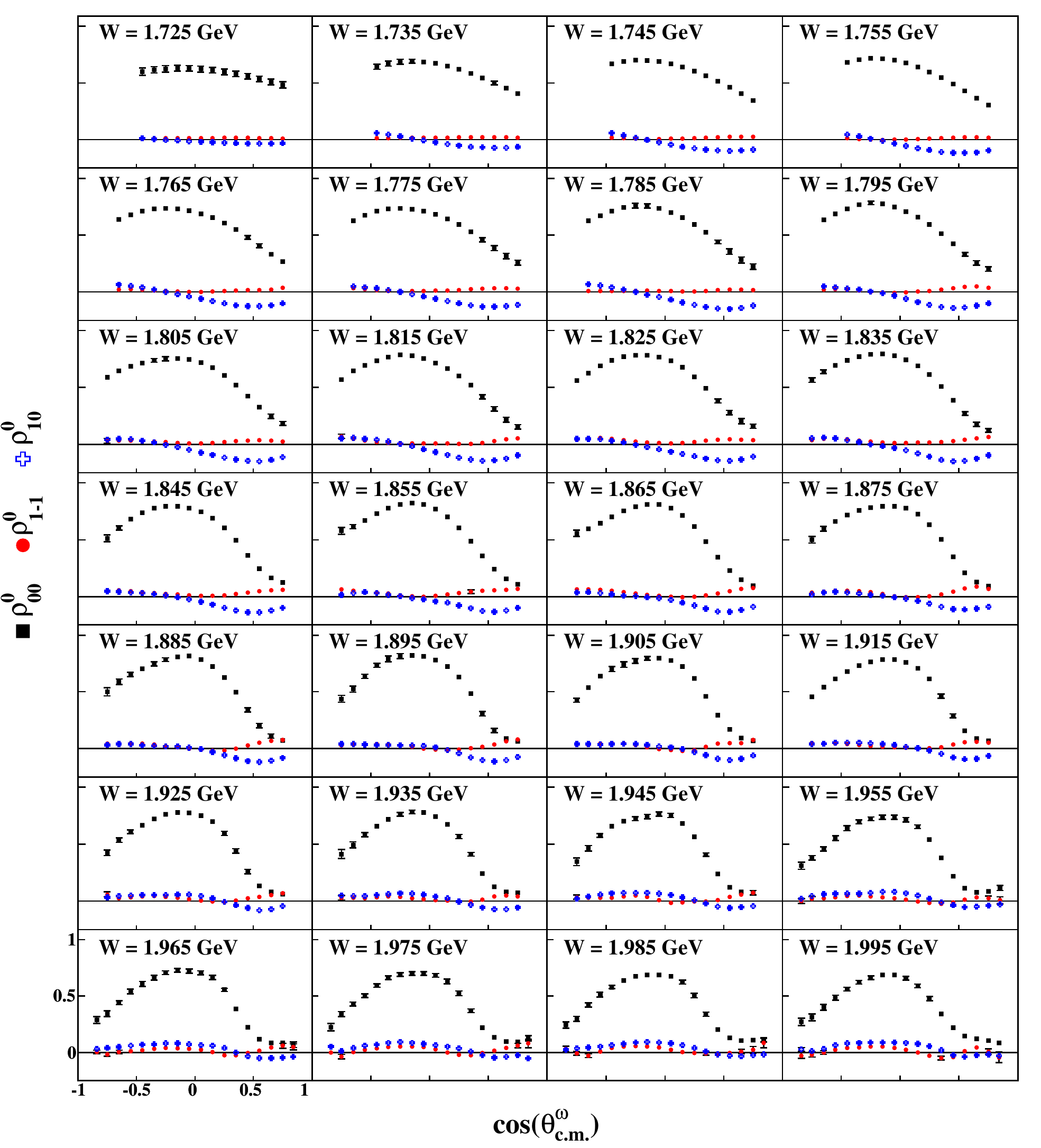}
\caption[]{\label{fig:rho-0}
  (Color Online)
  $\rho^0_{MM'}$ versus $\cos{\theta^{\omega}_{c.m.}}$: Spin density matrix 
  element measurements, in the Adair frame, 
  for bins in the range $1.72$~GeV $\leq W <$~$2.00$~GeV.
  The black squares 
  are $\rho^0_{00}$, the red circles are $\rho^0_{1-1}$ and the blue crosses
  are $Re(\rho^0_{10})$. 
  The centroid of each $10$-MeV wide bin is printed on the plot.
  The error bars do not include systematic uncertainties.
}
\end{figure*}

\begin{figure*}[p]
  \centering
 \hspace{-0.045\textwidth}
  \includegraphics[width=0.99\textwidth]{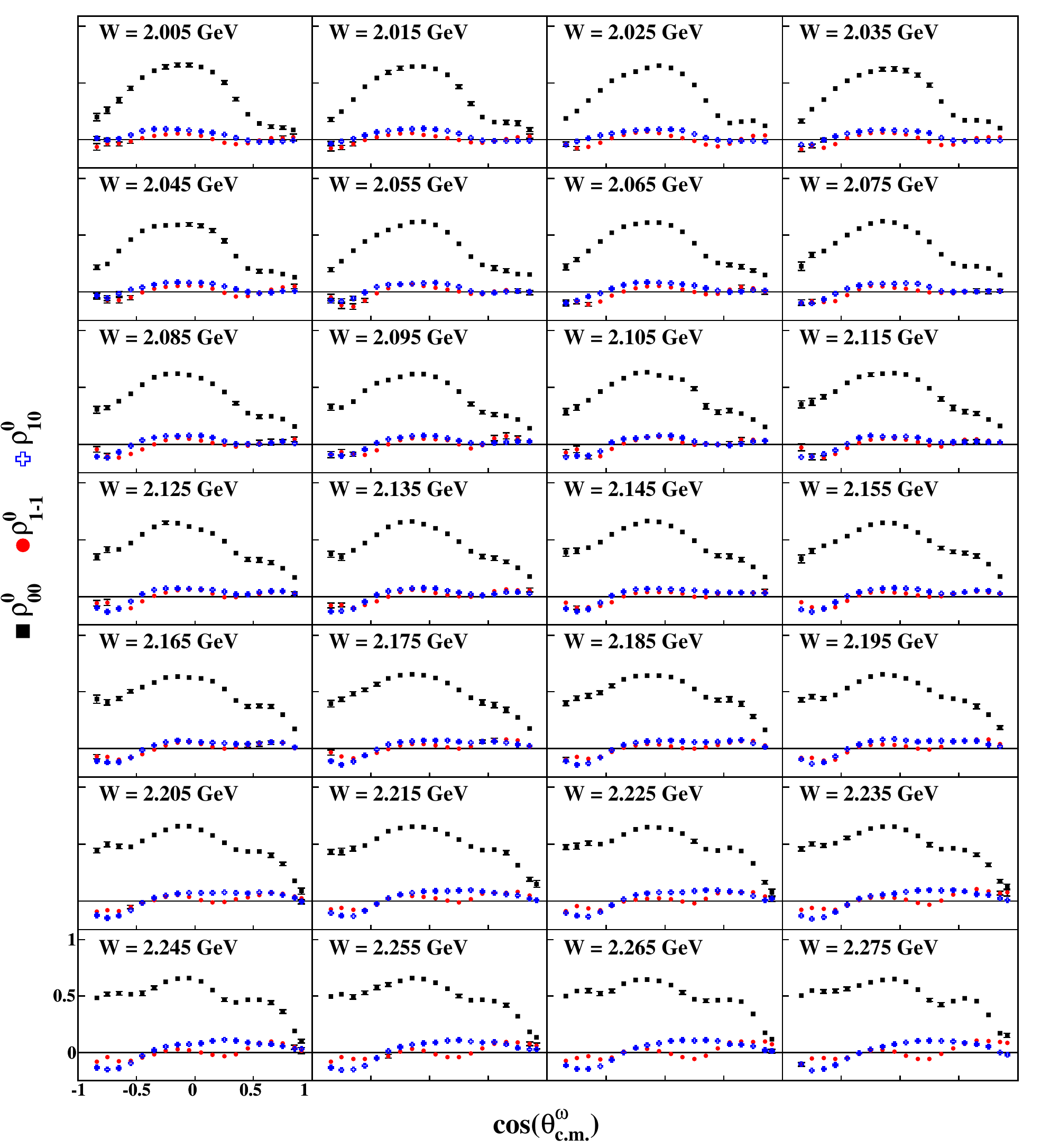}
\caption[]{\label{fig:rho-1}
  (Color Online)
  $\rho^0_{MM'}$ versus $\cos{\theta^{\omega}_{c.m.}}$: Spin density matrix 
  element 
  measurements, in the Adair frame, 
  for bins in the range $2.00$~ GeV $\leq W <$~$2.28$ GeV.
  The black squares 
  are $\rho^0_{00}$, the red circles are $\rho^0_{1-1}$ and the blue crosses
  are $Re(\rho^0_{10})$. 
  The centroid of each $10$-MeV wide bin is printed on the plot.
  The error bars do not include systematic uncertainties.
}
\end{figure*}

\begin{figure*}[p]
  \centering
 \hspace{-0.045\textwidth}
  \includegraphics[width=0.99\textwidth]{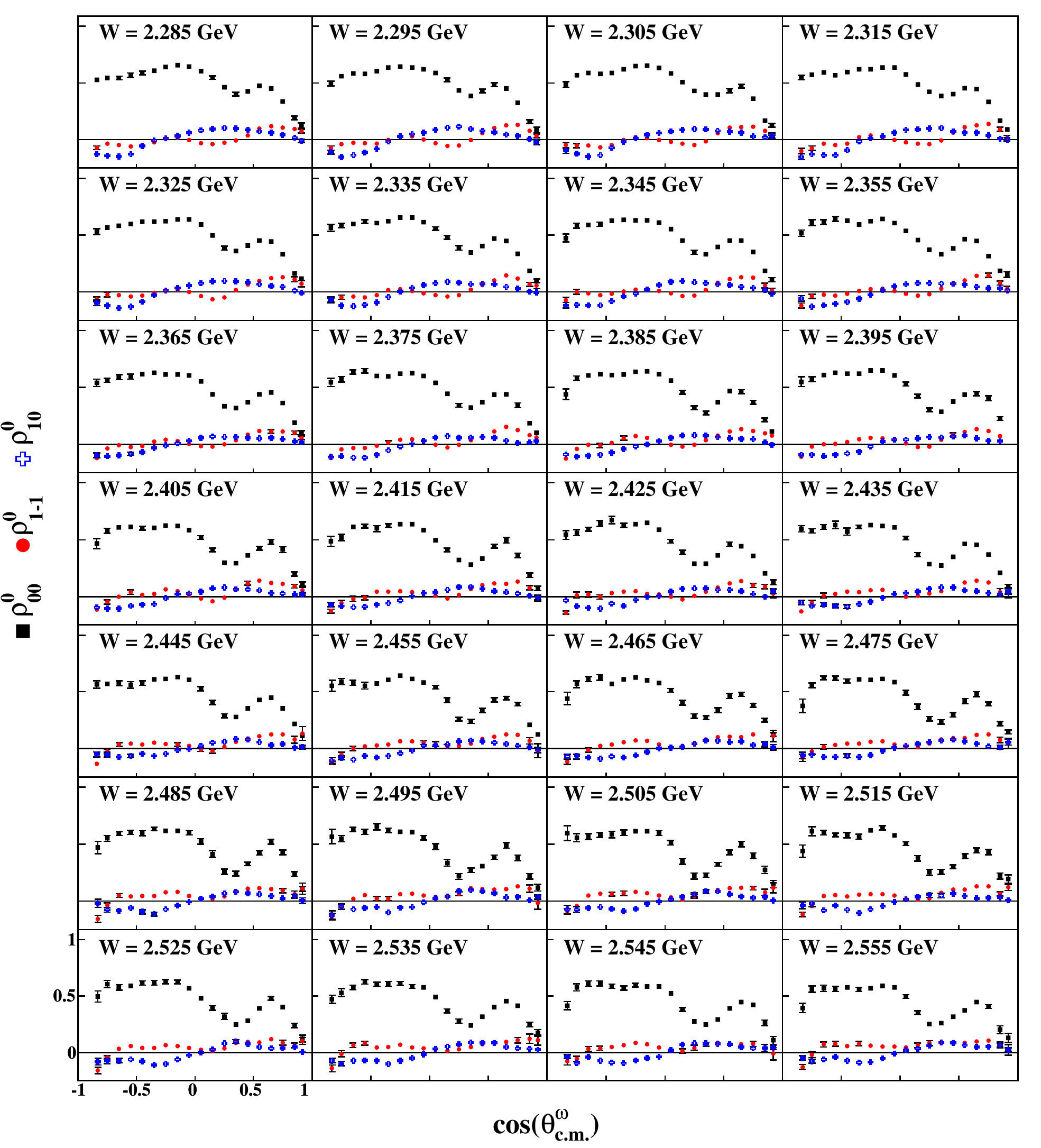}
\caption[]{\label{fig:rho-2}
  (Color Online)
  $\rho^0_{MM'}$ versus $\cos{\theta^{\omega}_{c.m.}}$: Spin density matrix 
  element 
  measurements, in the Adair frame, 
  for bins in the range $2.28$~GeV $\leq W <$~$2.56$ GeV.
  The black squares 
  are $\rho^0_{00}$, the red circles are $\rho^0_{1-1}$ and the blue crosses
  are $Re(\rho^0_{10})$. 
  The centroid of each $10$-MeV wide bin is printed on the plot.
  The error bars do not include systematic uncertainties.
}
\end{figure*}

\begin{figure*}[p]
  \centering
 \hspace{-0.045\textwidth}
  \includegraphics[width=0.99\textwidth]{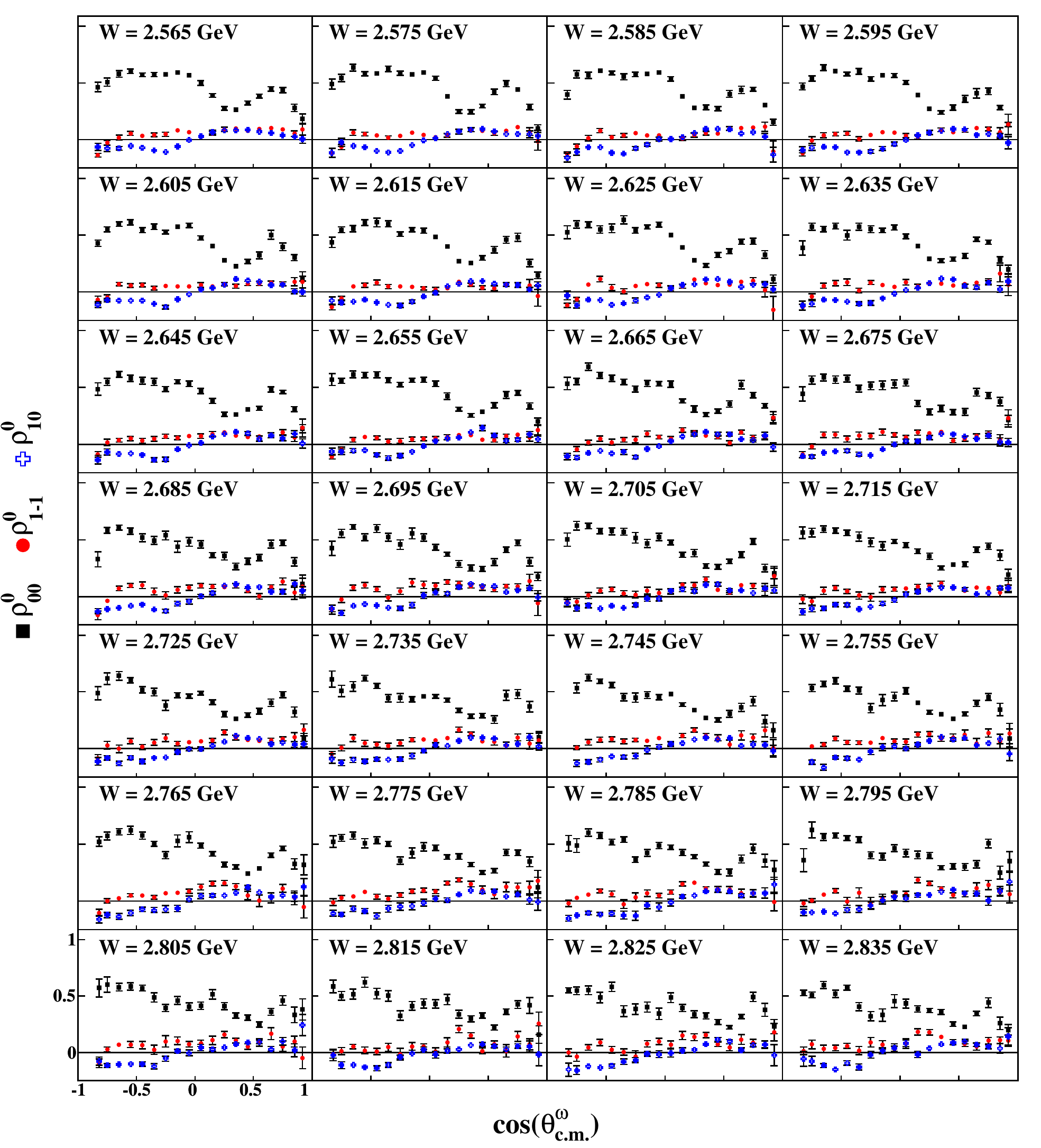}
\caption[]{\label{fig:rho-3}
  (Color Online)
  $\rho^0_{MM'}$ versus $\cos{\theta^{\omega}_{c.m.}}$: Spin density matrix 
  element 
  measurements, in the Adair frame, 
  for bins in the range $2.56$~GeV $\leq W <$~$2.84 GeV$.
  The black squares 
  are $\rho^0_{00}$, the red circles are $\rho^0_{1-1}$ and the blue crosses
  are $Re(\rho^0_{10})$. 
  The centroid of each $10$-MeV wide bin is printed on the plot.
  The error bars do not include systematic uncertainties.
}
\end{figure*}

Figs.~\ref{fig:rho-0}, \ref{fig:rho-1}, \ref{fig:rho-2} and \ref{fig:rho-3} show the $\rho^0_{MM'}$ results
extracted using the partial wave expansion technique. 
The error bars are purely statistical. 
The spin density matrix elements do not rely on normalization information;
thus, only the acceptance can contribute to the systematic uncertainty.
Possible effects due to systematic problems in the acceptance calculation were
examined by analyzing decay distributions distorted by our estimated acceptance
uncertainty. Based on this study, we estimate the systematic uncertainties in
our results to be as follows:
\begin{subequations}
\begin{eqnarray}
  \sigma_{00} &=& 0.0175 \\
  \sigma_{1-1} &=& 0.0125\\
  \sigma_{10} &=& 0.01.
\end{eqnarray}
\end{subequations}

Over most of our kinematics, these results represent the first high-precision
measurements of $\rho^0_{MM'}$ for $\omega$ photoproduction. 
Near threshold and at forward angles, the cross section develops a strong 
forward peak, which is indicative of $t$-channel contributions. 
In this same region, 
the diagonal $\rho^0_{00}$ element decreases sharply as
the energy increases, or equivalently, as the forward peak increases in 
significance. This is typical of exchange of a spin-0 particle in the 
$t$-channel where the 
$\omega$ is forced to carry the spin of the photon at forward angles. 
This new precise polarization information should help determine the relative
strengths of the scalar and pseudoscalar exchanges 
(see Sec.~\ref{sec:interpretation}).

At higher energies, starting near $W \sim 2.1$~GeV, a dip in 
$\rho^0_{00}$ appears at $\cos{\theta^{\omega}_{c.m.}} \sim 0.4$, which continues
to increase in prominence until about $W \sim 2.5$ GeV. 
Above this energy, its significance slowly decreases; however, it is still 
present at our highest energies. 
This dip is located near where the forward peak (typically associated with 
$t$-channel contributions) has decreased in significance such that it is 
approximately the same size as the cross section in the region 
$0 < \cos{\theta^{\omega}_{c.m.}} < 0.4$. 
Thus, it is possible that this dip results from interference between the 
$t$-channel and larger-angle production mechanisms. 
In the kinematic regions where the cross section possesses the humps and dips discussed 
in Sec.~\ref{section:results:dsigma}, there are a number of interesting features found in the 
spin density matrix elements as well. The partial wave analysis we performed
on this data found that these features are well described by  
baryon resonance contributions~\cite{williams-prd}. 

\subsection{\label{sec:interpretation}Interpretation of the data}
In the low-energy regime, these new measurements have been used to carry out a mass-independent 
partial wave analysis of the reaction $\gamma p\rightarrow \omega p$. The results of this analysis, 
which are presented in a concurrent article~\cite{williams-prd} and are not discussed in detail here, show clear evidence of $s$-channel 
resonance contributions. 
This PWA, the results of which are different from previous 
analyses~\cite{paris-09,zhao-2001,titov,oh-2001,penner-2002,penner-2005},
was the first to benefit from the strong
additional constraints provided by the high-precision polarization results
obtained from these data.

The high-energy measurements have been compared to two existing models for 
$\omega$ photoproduction.  The first is the model of 
Oh {\em et al}~\cite{oh-2001} which includes pseudoscalar meson 
($\pi^{0}$ and $\eta$) and Pomeron exchange in the $t$-channel, along with 
nucleon exchange in both the $s$- and $u$-channel. It also includes 
$s$-channel contributions, which are necessary to describe the data in the 
central region of the angular range. The second model is that of 
Laget~\cite{laget-2000,laget-2002} which includes $t$- and $u$-channel
contributions similar to that of Ref.~\cite{oh-2001}, but also allows for a 
contribution from two-gluon exchange. In this latter model, the two-gluon term 
is required to describe the $\phi$ photoproduction data.

\begin{figure}[h!]\centering
\includegraphics[width=0.53\textwidth]{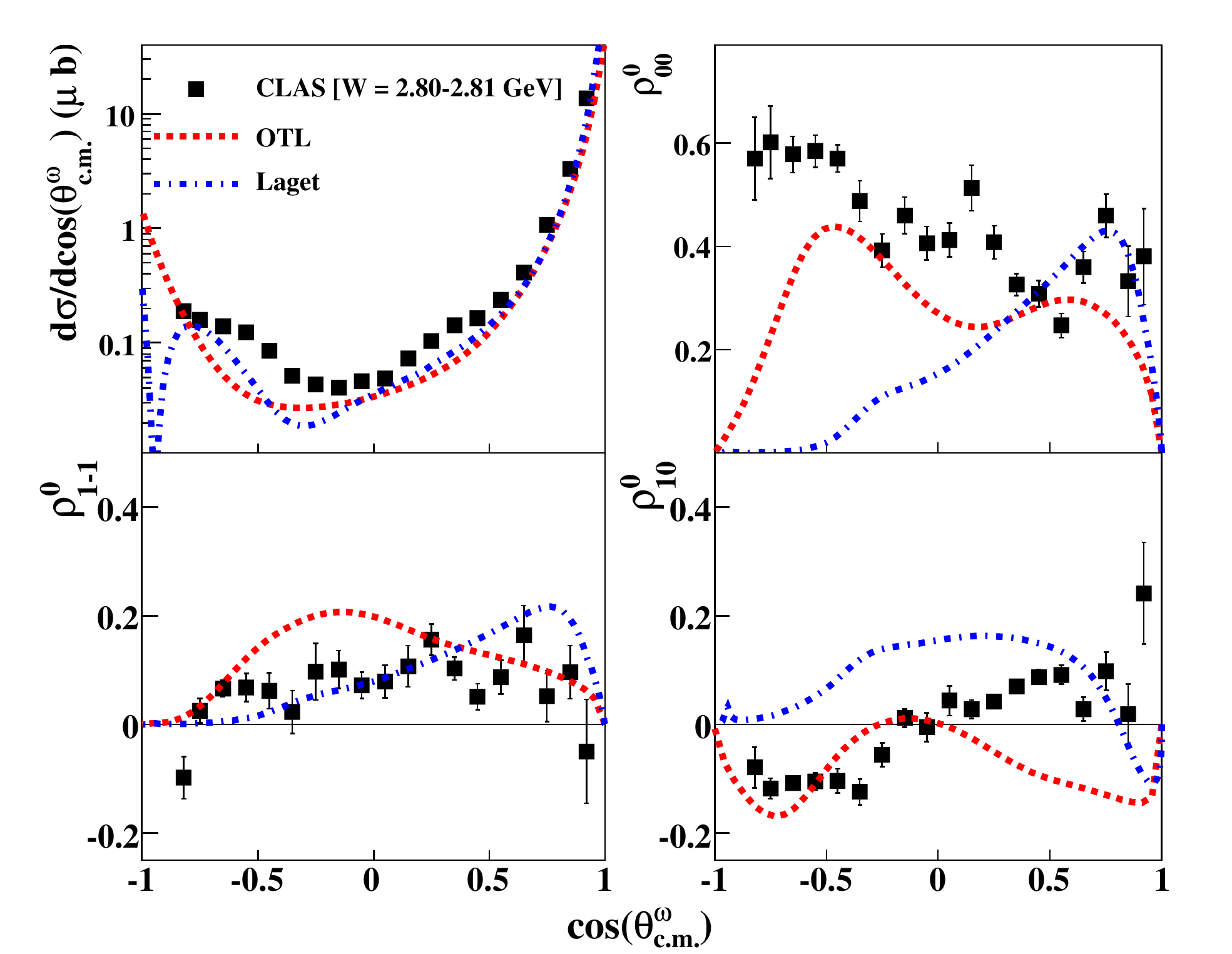}
\caption[]{\label{fig:t-compare}
  A comparison of the theoretical models of Oh {\em et al}~\cite{oh-2001} 
  (red dashed lines) and Laget~\cite{laget-2000,laget-2002} (blue dot-dashed
  lines) to the $W=2.805$~GeV data presented in this article (black squares).
}
\end{figure}

Fig.~\ref{fig:t-compare} shows comparisons of these models to our
data at $W=2.8$~GeV.  Both models do a reasonable job of reproducing the 
cross-section measurements. The $t$-channel terms drive the very forward-angle 
data where the agreement is very good. At backwards angles, where the 
$u$-channel terms dominate, the agreement is not as good as it is at forward 
angles. In the central region, both models agree with the overall shape of
the cross section; however, the finer structure in the data is not reproduced.

Neither model is able to reproduce the new high-precision spin density matrix
element measurements presented in this article. While some regions are 
reasonably well described by one model or the other, neither gives anything 
close to good overall agreement. Perhaps the most striking discrepancy is that 
at forward angles, where the cross sections are described very well by both
models, neither provides an excellent description of the spin density 
matrix elements. The high-precision measurements presented in
this article clearly provide new stringent constraints on both the nature of 
the production mechanisms in the high-energy regime, as well as on the search 
for missing baryon resonances.

\subsection{Comparison to previous measurements}
Previous experimental measurements that overlap our energy range have been 
made at CLAS in 2003~\cite{battaglieri-2003}, at SAPHIR in 
2003~\cite{saphir-2003}, at Daresbury in 1984~\cite{barber-1984} and
1977~\cite{clift-1977}, and at SLAC in 1973~\cite{ballam-1973}. Below we 
compare our measurements with each of these previous results. The
cross sections will be examined first, 
followed by the spin density matrix elements.

\begin{figure}
  \centering
  \includegraphics[width=0.49\textwidth]{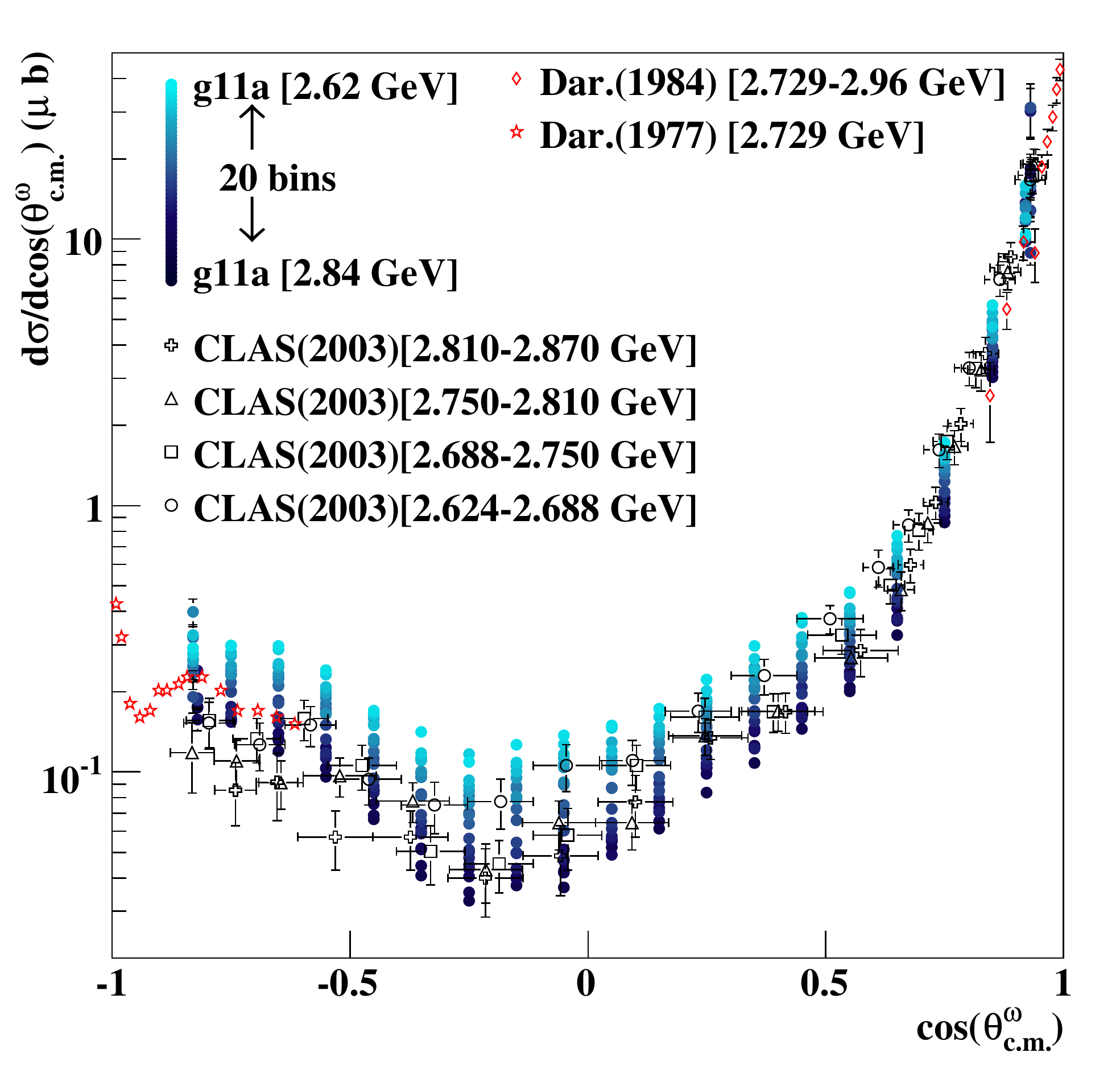}
\caption[]{\label{fig:dsigma-compare-high-w}
  (Color Online)
  $\frac{d\sigma}{d\cos{\theta^{\omega}_{c.m.}}} (\mu b)$ versus 
  $\cos{\theta^{\omega}_{c.m.}}$:
  Comparison of the present CLAS results (blue circles) with previously 
  published results from CLAS~\cite{battaglieri-2003} 
  (black open crosses, triangles, squares and circles) and 
  Daresbury~\cite{clift-1977,barber-1984} (red open diamonds and stars) 
  in the energy range 2.62~GeV~$<~W~<~$~2.96~GeV. The Daresbury (1977)
  points have no error bars; the points were extracted from a portable document
  format (PDF) image.
}
\end{figure}

\begin{figure*}[p]
  \begin{center}
  \includegraphics[width=0.8\textwidth]{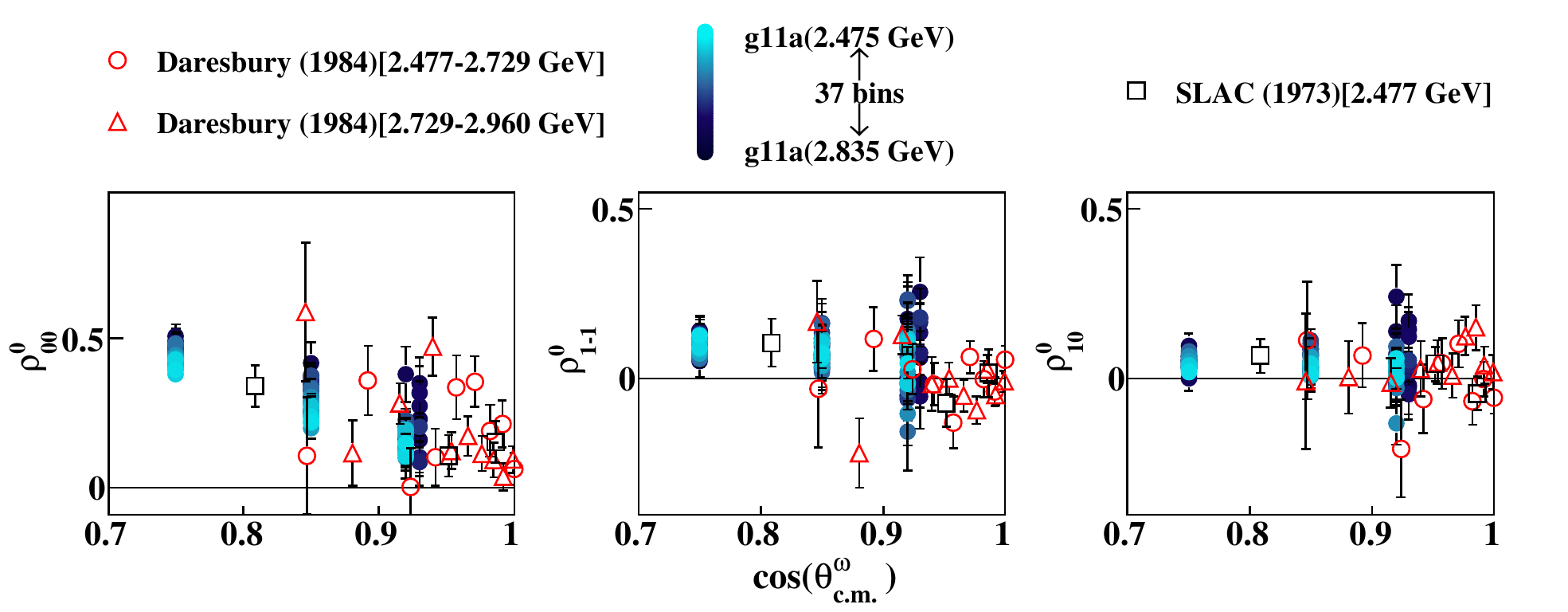}
  \caption[]{\label{fig:rho-compare-high-w}
    (Color Online)
    $\rho^0_{MM'}$ in the Adair frame versus $\cos{\theta^{\omega}_{c.m.}}$: 
    Comparison of the present CLAS results (blue circles) with previously
    published Daresbury~\cite{barber-1984} (open red circles and triangles)
    and SLAC~\cite{ballam-1973} (open black squares).
  }
  \end{center}
\end{figure*}

Fig.~\ref{fig:dsigma-compare-high-w} shows a comparison of the cross-section
results presented in this article with previously published results from 
CLAS~\cite{battaglieri-2003} and Daresbury~\cite{clift-1977,barber-1984}. 
The previous CLAS 
results, four energy bins in the range 2.624~GeV~${<W<}$~2.87~GeV,
cover virtually the same angular range as the current results. The agreement
is very good for $\cos{\theta^{\omega}_{c.m.}} > -0.1$; however, there is a
sizable discrepancy in the backward direction. At the time of the earlier
CLAS measurement, the $\omega$ polarization had only been measured in the 
forward direction (see Fig.~\ref{fig:rho-compare-high-w}); 
thus, these values of the
spin density matrix elements were used in the acceptance calculation. Our
results show that the polarization is quite different at backward and forward
angles. Near the edges of the CLAS acceptance, {\em e.g.} in the backward 
direction,
an incorrect description of the polarization can lead to large errors
in the acceptance calculation. This is most likely the cause of the 
discrepancy in the cross sections.
The Daresbury results, which were only published in the very forward and 
backward regions, are in good agreement with our measurements.

\begin{figure*}[p]
  \centering
  \hspace{-0.045\textwidth}
  \includegraphics[width=0.99\textwidth]{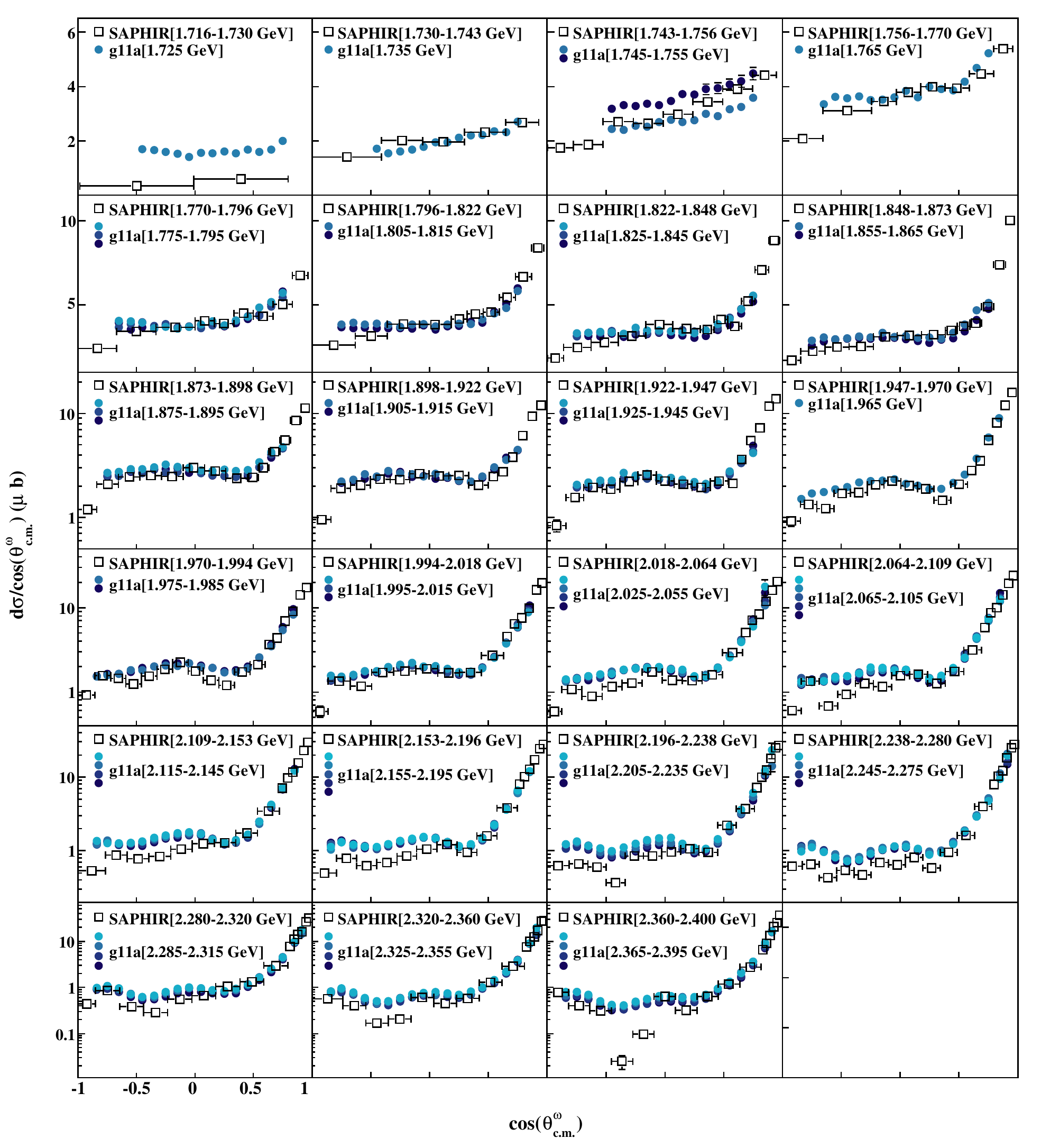}
\caption[]{\label{fig:dsigma-compare-saphir}
  (Color Online)
  $\frac{d\sigma}{d\cos{\theta^{\omega}_{c.m.}}} (\mu b)$ versus
  $\cos{\theta^{\omega}_{c.m.}}$:
  Comparison of the present CLAS results (blue circles) with previously 
  published results from SAPHIR~\cite{saphir-2003} (black open squares). 
}
\end{figure*}

For $W < 2.4$~GeV, the previous large acceptance results come from 
SAPHIR~\cite{saphir-2003}. Fig.~\ref{fig:dsigma-compare-saphir} shows a 
comparison of the SAPHIR cross-section measurements with the present CLAS 
results. The error bars shown for the SAPHIR points do not include systematic 
uncertainties.
The agreement is fair, but there are some discrepancies. The SAPHIR experiment 
had better angular coverage; however, the CLAS results are more precise. 
In the forward direction, the agreement is very good at all energies. 
At moderate angles, $|\cos{\theta^{\omega}_{c.m.}}| < 0.5$, the agreement is 
good at lower energies but the CLAS results tend to be higher as the energy 
increases. In the backward direction, where the CLAS has acceptance, 
the CLAS points are almost always higher than the SAPHIR points.

\begin{figure*}[p]
  \begin{center}
  \includegraphics[width=0.8\textwidth]{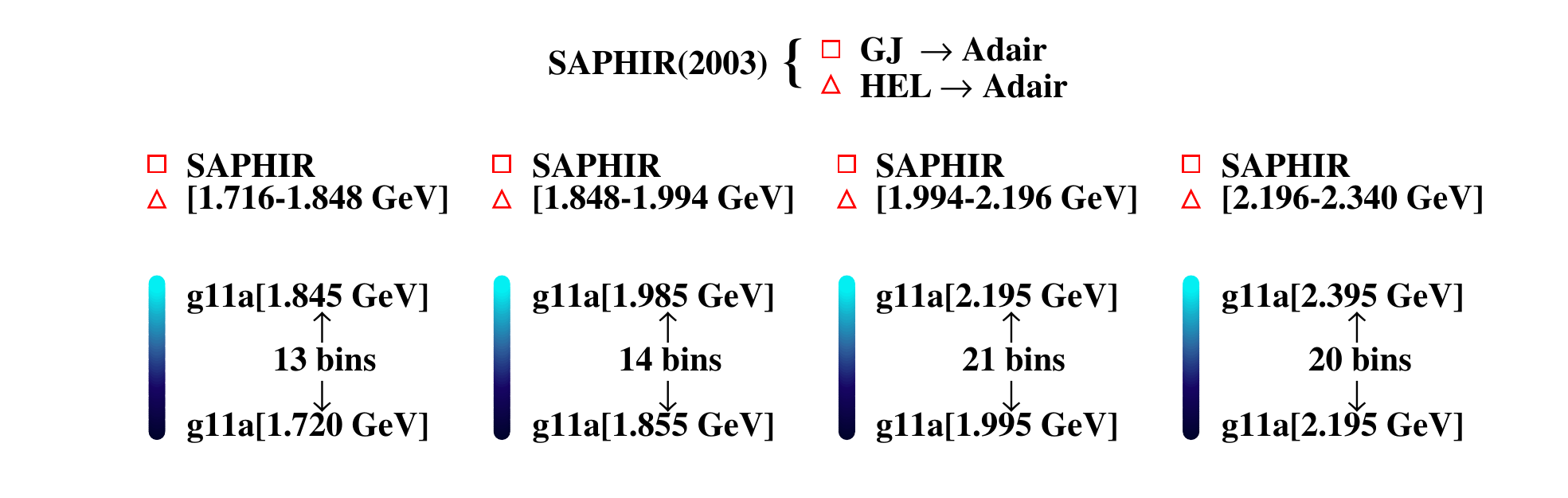}\\
  \includegraphics[width=0.8\textwidth]{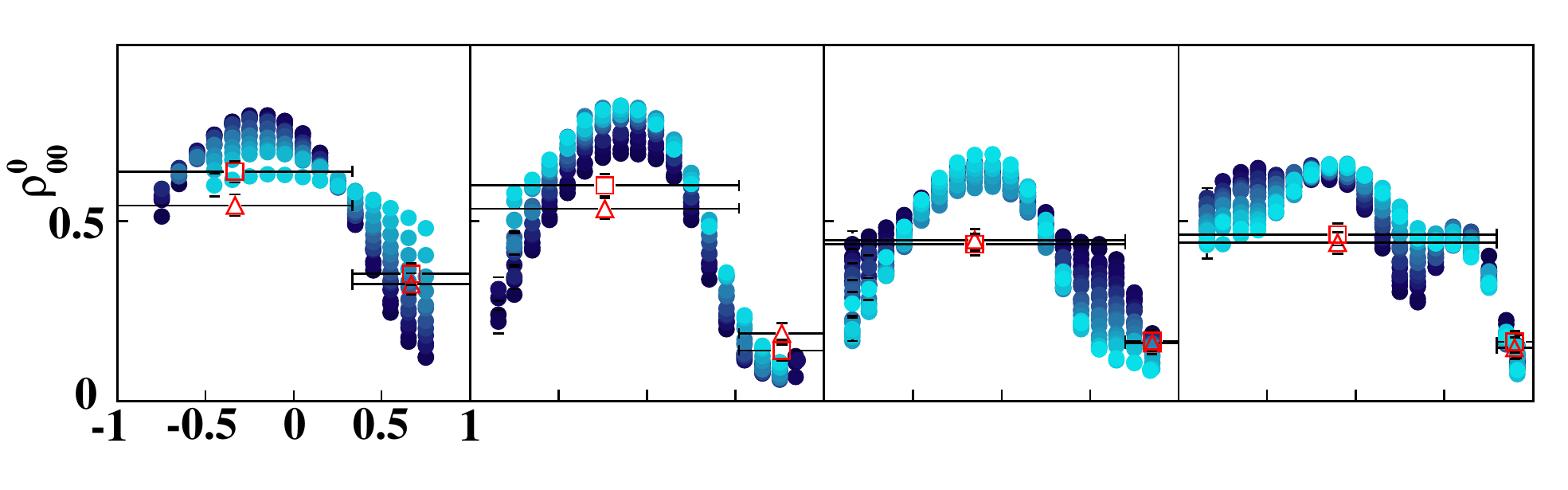}\\
  \includegraphics[width=0.8\textwidth]{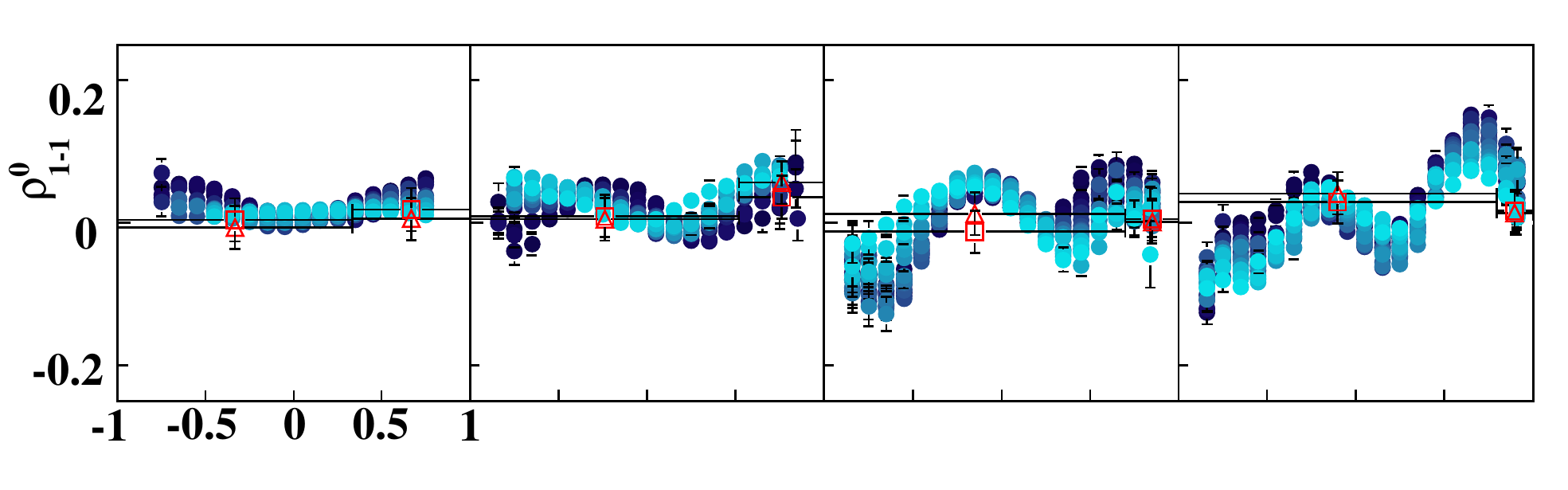}\\
  \includegraphics[width=0.8\textwidth]{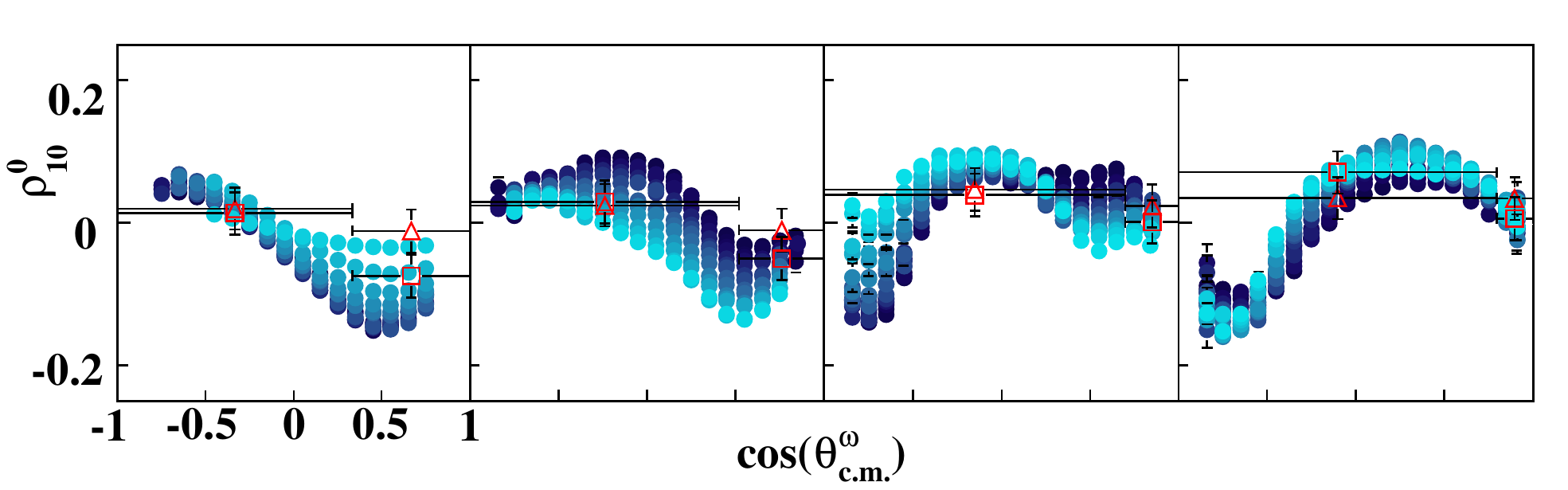}
  \caption[]{\label{fig:rho-compare-saphir}
    (Color Online)
    $\rho^0_{MM'}$ in the Adair frame versus $\cos{\theta^{\omega}_{c.m.}}$: 
    Comparison of the present CLAS results (blue circles) with previously
    published SAPHIR~\cite{saphir-2003} results (open red squares and 
    triangles).
    SAPHIR extracted results independently in the Gottfried-Jackson and
    Helicity frames --- both presented here rotated to the Adair frame.
  }
  \end{center}
\end{figure*}

Previous spin density matrix element measurements are much less precise. 
The only results published for $W < 2.4$~GeV come from 
SAPHIR~\cite{saphir-2003}. Fig.~\ref{fig:rho-compare-saphir} shows a 
comparison
of the SAPHIR results, which consist only of four energy bins, each with two
angular points, and the present CLAS results, which include 1371 total data 
points in this energy range. 
We note here that the SAPHIR collaboration published their results in 
both the Gottfried-Jackson and Helicity frames, with each measurement 
constituting an independent fit to their data. 
Both results were rotated into the Adair frame for comparison.
Overall, the SAPHIR results are in good agreement with our measurements.


At higher energies, previously published results only exist at very forward
angles. Fig.~\ref{fig:rho-compare-high-w} shows a comparison of our forward
high energy results with those from Daresbury~\cite{barber-1984} and
SLAC~\cite{ballam-1973}. The agreement is good. For $W > 2.4$~GeV, the
results presented in this article for $\rho^0_{MM'}$ are the world's first 
measurements for $\cos{\theta^{\omega}_{c.m.}} < 0.8$.
\section{\label{section:conc}Conclusions}

In summary, experimental results for $\omega$ photoproduction from the proton
have been presented in the energy regime from threshold up to 
$W = 2.84$~GeV. Both differential cross section and spin density matrix
element measurements are reported. The cross-section results are
the most precise to date and provide the largest energy and angular coverage. 
The results are in fair to good agreement with previous experiments. 
For $W < 2.4$~GeV, we present 1181 $\rho^0_{MM'}$ data points; 
the previous world's data consisted of 8 points. At higher energies, we have
made the first spin density matrix element measurements for 
$\cos{\theta^{\omega}_{c.m.}} < 0.8$. Our $\rho^0_{MM'}$ measurements are in 
good agreement with the, rather sparse, existing data. The 1960 
$(W,\cos{\theta^{\omega}_{c.m.}})$ cross-section
points, along with the 2015 $(W,\cos{\theta^{\omega}_{c.m.}})$ 
spin density matrix element data points can be obtained at Ref.~\cite{cite:clas-db}.

These new data will have a large impact on our current understanding of 
vector-meson photoproduction, as well as provide a crucial data set in 
the search for missing baryon resonances. 
A mass-independent partial wave analysis performed on these data, 
which is the
first such analysis to benefit from the strong constraints provided by 
high-precision polarization information, found strong evidence for baryon
resonance contributions~\cite{williams-prd}. Furthermore, none of 
the current models of high-energy $\omega$ photoproduction are able to describe
the precise spin density matrix element measurements presented in this article.
We look forward to seeing what impact these new results will have on future
models of vector-meson photoproduction.

\begin{acknowledgments}
We thank the staff of the Accelerator and the Physics Divisions at
Thomas Jefferson National Accelerator Facility who made this
experiment possible.  
This work was supported in part by the U.S. Department of Energy
(under grant No. DE-FG02-87ER40315), 
the National Science Foundation,
the Italian Istituto Nazionale di Fisica Nucleare, 
the French Centre National de la Recherche Scientifique, 
the French Commissariat \`{a} l'Energie Atomique, 
the Science and Technology Facilities Council (STFC),
and the Korean Science and Engineering Foundation.  
The Southeastern Universities Research Association (SURA)
operated Jefferson Lab under United States DOE contract
DE-AC05-84ER40150 during this work.
\end{acknowledgments}

\end{document}